\documentclass[final,5p,times,twocolumn,sort&compress]{elsarticle}

\usepackage{amssymb}
\usepackage{lipsum}
\usepackage{mathtools}
\usepackage{graphicx}
\usepackage{braket}
\usepackage{caption}
\usepackage{subcaption}

\usepackage{slashed}
\journal{Physics Letters B}
 
\begin{document}

\begin{frontmatter}

\title{Persistent homology of collider observations: when (w)hole matters}
\author[first,second]{Jyotiranjan Beuria}
\affiliation[first]{organization={IKSMHA Centre, Indian Institute of Technology, Mandi},country={India}}
\affiliation[second]{organization={Indian Knowledge Systems Centre, ISS Delhi},country={India}}

\begin{abstract}
Topological invariants have played a fundamental role in the advancement of theoretical high energy physics. Physicists have used several kinematic techniques to distinguish new physics predictions from the Standard Model (SM) of particle physics at Large Hadron Collider (LHC). However, the study of global topological invariants of the collider signals has not yet attracted much attention. In this article, we present a novel approach to study collider signals using persistent homology. The global topological properties of the ensemble of events as expressed by measures like persistent entropy, Betti area, etc. are worth considering in addition to the traditional approach of using kinematic variables event by event. In this exploratory study, we first explore the characteristic topological signature of a few SM electroweak resonant productions. Next, we use the framework to distinguish global properties of the invisible Higgs decay processes in the SM and a real singlet extension of the SM featuring stable singlet scalar dark matter.
\end{abstract}



\begin{keyword}
%% keywords here, in the form: keyword \sep keyword, up to a maximum of 6 keywords
Persistent Homology \sep Topological Data Analysis \sep LHC \sep BSM Physics



\end{keyword}

\end{frontmatter}

%\tableofcontents



\section{Introduction}
\label{introduction}
The Standard Model (SM) of particle physics has been immensely successful in describing the dynamics of elementary particles. The discovery of a neutral scalar particle, Higgs boson \citep{aad2012observation,chatrchyan2012observation} was indeed the triumph of the SM. Despite the remarkable success of the SM, it fails to answer several essential questions. Thus, the quest for a new physics model beyond the SM (BSM) has led to several phases of upgrade for the Large Hardon Collider (LHC). With the third phase run of the LHC gathering data, it is imperative to devise new methods to discriminate a plethora of BSM models.

The study of the phenomenology of BSM models involves constraining the parameter space of new physics models with collider simulation at the patron level, hadron level and detector level through signal and background analysis. Physicists have devised several kinematic variables \cite{Franceschini:2022vck} for this matter. This approach relies on kinematic cuts applied on event by event basis. 
There have been some attempts to study the global properties of events using Voronoi and Delaunay tessellations \cite{Debnath:2015wra, Debnath:2016mwb, Matchev:2020vhr}, network distance metrics \cite{Mullin:2019mmh} and multi-event ML classifier \cite{Nachman:2021yvi}, etc.  However, Topological Data Analysis (TDA) \cite{taylor2015topological,topaz2015topological,lloyd2016quantum,gidea2018topological, saggar2018towards,sizemore2019importance, Murugan:2019alb, cole2019topological, chazal2021introduction} for studying the global properties of events have not attracted much attention.

% To the best of our knowledge, we are unaware of any discussion on the global properties of the signal or background space wherein the entirety of the signal or background is discussed.

Algebraic topology is a branch of mathematics that studies the global properties of topological spaces in terms of the associated invariants. TDA has attracted much interest in recent years in the broader field of data science as a complement to more traditional machine learning methods. Of all such computational geometry techniques, persistent homology stands out in terms of its ease of application and predictive power. Homology is all about the $k-$dimensional holes that are the properties of the system as a whole. The applications of persistent homology range from multi-dimensional biological networks to social networks. It is a tool to analyze the topological characteristics of data that persist over various scales of filtration. 

Information networks are crucial tools for simulating the dynamics associated with the relations among the components of a complex system. The traditional approach using network graphs is limited by binary relations. However, relations modeled through simplicial complexes have much richer phenomenology because of the possible multi-dimensional relationships between different parts of the system. 

Interestingly, simplicial complexes form a topological space. Thus, the associated dynamics is characterized by topological invariants. The fundamental idea behind persistent homology is to substitute data points with a parametrized family of simplicial complexes, which are roughly thought of as a union of points, edges, triangles, tetrahedrons, and other higher-dimensional polytopes. Such a topological space encodes the change in the topological features (such as the number of connected components, holes, voids, etc.) of the simplicial complexes across different parameters.

The dynamics involved in particle physics colliders is highly complicated. With run-3 of the LHC in progress, there is a great need to understand the features of the data collected at the LHC. The BSM physics models are likely to leave behind very complicated relations among the detected particles, characterizing the specific model under discussion. There have been several extensions of the SM proposed, viz., Supersymmetry \cite{csaki1996minimal,martin1998supersymmetry,Baer:2006rs,ellwanger2010next}, Minimal Universal Extra Dimension \cite{cembranos2007exotic, datta2010minimal}, etc. are a few popular ones to name. As an initial study to demonstrate the global properties of data, we explore TDA techniques for the SM and a real scalar extension of the SM. The real singlet scalar extension of the SM is a simple model with rich Higgs phenomenology and has attracted much attention \cite{ham2005electroweak,barger2008cern,guo2010real,enqvist2014standard,feng2015closing,kanemura2016radiative,kurup2017dynamics,chiang2019revisiting}. Since many BSM models feature real singlet scalars, this model is also well motivated as a low energy effective model. In this work, we focus on a particular variant wherein the singlet scalar becomes the stable particle under a global $Z_2$ symmetry \cite{guo2010real,feng2015closing}, thus serving as a candidate for dark matter. This model is sometimes dubbed the Singlet Scalar Dark-Matter Model (SSDM). The framework of analysis using persistent homology is quite generic and can be used for any other model as well. We choose this simple model to demonstrate the usability of persistent homology in search for new physics.

The organization of the paper is as follows. In section \ref{sec:ph}, we give a preliminary introduction to the mathematics of persistent homology. In section \ref{sec:framework}, we discuss the framework of topological analysis and collider simulation. In section \ref{sec:ph-sm}, we compare the global topological signatures associated with various SM processes. In section \ref{sec:ssdm}, we study the parameter space of the SSDM and subject it to the constraints from the measurements of the 125 GeV Higgs boson at colliders. In section \ref{sec:ph-ssdm}, we compare the persistent homology of the invisible decay of the Higgs boson in the SM with the SSDM. In section \ref{sec:conclusion}, we summarize and conclude the discussion.

\section{Simplicial Complex and Persistent Homology}
\label{sec:ph}
% \lipsum[1]

Simplicial complex is the fundamental geometrical structure associated with the study of persistent homology. This section offers some background notions of simplicial complex and persistent homology. We also discuss the topological properties we will extensively use in our analysis. Our reader is advised to refer to some excellent reviews \cite{carlsson2009topology,Murugan:2019alb,sizemore2019importance,chazal2021introduction} on Topological Data Analysis (TDA) available on the web. One can also refer to several worked out real-life examples \cite{Giotto:2023,Gudhi:2023} with code.

\subsection{Simplicial complexes}
\begin{figure}[t]
\centering 
\includegraphics[width=0.45\textwidth, angle=0]{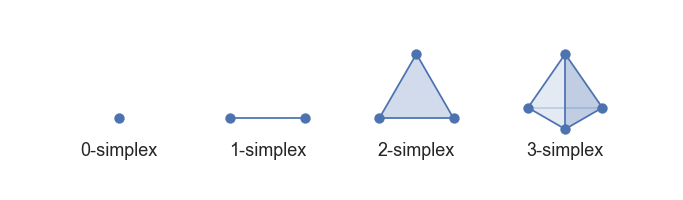}	
\caption{Simplices (plural of simplex) are the combinatorial building blocks of a simplicial complex. For illustration, 0-,1-,2-, and 3-simplex are shown from left to right.} 
\label{fig:simplex}%
\end{figure}

Simplicial complexes are a way to build topological spaces from basic combinatorial building blocks. It reduces the complexities of the continuous geometry of topological spaces and instead deals with the task of comparatively simple combinatorics and counting. These simple combinatorial building blocks are called simplices, as illustrated in figure \ref{fig:simplex}. A $k-$dimensional simplex is formed by taking the convex hull of $k+1$ independent points. Thus, 0-simplex is a point, 1-simplex is a line segment, 2-simplex is a filled triangle and 3-simplex is a filled tetrahedron.
Similarly, one can construct higher-dimensional polytopes. For a $n-$dimensional simplex, the simplices with dimension $k<n$ form the faces. Thus, for a 2-simplex (triangle), its edges (1-simplex) are the faces.

A simplicial complex ($K$) is a finite collection of simplices ($\sigma$) such that for every $\sigma \in K$,
\begin{enumerate}
  \item any face of $\sigma \in K$ is also part of $K$ and
  \item if $\sigma_1, \sigma_2 \in K$, then $\sigma_1 \cap \sigma_2 $ is a face of both $\sigma_1$ and $\sigma_2$.
\end{enumerate}
A collection of $n-$dimensional data points is typically represented as point cloud data (PCD). Even a time series data can also be embedded as point cloud data using Taken's embedding theorem \cite{takens2006detecting}. There are several algorithms to convert point cloud data ($X$) to simplicial complex. For our purpose, we will use Vietoris–Rips complex algorithm \cite{vietoris1927hoheren}, which is one of the simplest and computationally less expensive.

Let $X = \{x_0,x_1,...,x^{m-1}\} \in R^{m}$ be the point cloud data with each point $x_i$ in $R^{n}$. Let $r$ be a fixed radius. The Vietoris-Rips complex of $X$ is an abstract simplicial complex whose $k-$simplices are the (k + 1)-tuples of points in $X$ such that the pairwise distance between them is less than equal to $r$. This maximal radial limit $r$ is also termed the filtration parameter. Mathematically, the Vietoris-Rips complex (also known as Rips complex) is given by

\begin{equation}
VR_r(X)=\{\sigma \subseteq X \: | \:  d(x_i,x_j) \leq r \; \;\forall x_i \neq x_j \in \sigma\},
\label{eq:rips}
\end{equation}
where $d(x_i,x_j)$ is the Euclidean distance between two points. It is to be noted that a metric other than Euclidean distance can also be taken.
\subsection{Chain complexes and Homology groups}
A $k-$chain denoted by $C_k$ is the formal sum of a subset of $k-$simplices of the simplicial complex $K$. $C_k$ can be expressed as 
\begin{equation}
  C_k=\sum_{i} \alpha_i \sigma_k^i \;,
\end{equation}
where $\sigma_k$ is the $k-$simplex and $\alpha_i$ is assumed to be a real number, i.e., $\alpha_i \in R$. It is interesting to note that $C_k$ forms an abelian group under component-wise addition operation. The $k-$chain group is also generated by the $k$-cycles, where $k$-cycle is a $k-$chain without boundary. The $k-$th boundary operator on $k-$simplex $\sigma_k$ with vertices $(v_0, v_1, ...,v_k)$ is expressed as 
\begin{equation}
    \partial_k(\sigma_k)=\sum_{i=0}^{k} (-1)^i (v_0, v_1, ..., \hat{v}_i, ..., v_k),
\end{equation}
where $\hat{v}_i$ is the vertex deleted from $\sigma_k$. In other words, we have
\begin{equation}
  \partial_k:\; C_k \rightarrow C_{k-1}\;,
\end{equation}
where $k-$chain $C_k$ is mapped to $(k\!-\!1)$-chain $C_{k-1}$ under boundary operation. A chain complex is created by the collection of boundary operators on the chain groups,
\begin{equation}
C_k \xrightarrow{\partial_k} C_{k-1} \xrightarrow{\partial_{k-1}} C_{k-2}  \;\;... \;\; C_1 \xrightarrow{\partial_1} C_0  \xrightarrow{\partial_0} 0
\end{equation}

The kernel of the boundary operator $\partial_k$ is the set of all $C_k$ that has no boundary and the image of $\partial_k$ is 
the set of $(k\!-\!1)$-chains $C_{k-1}$ that are the boundaries of $C_k$. This can be expressed mathematically as 
\begin{gather} 
ker({\partial_k})=\{ c \in C_k \;|\; \partial_k {C_k}=0  \} \nonumber \\
im({\partial_k})=\{ d \in C_{k-1} \;|\;  \exists c \in C_k : d=\partial_k ({c})  \}
\end{gather}
Thus, we find that elements of $ker({\partial_k})$ are nothing but $k-$cycles while $k-$boundary is an element of $im({\partial_{k+1}})$. It is interesting to note that the sets of $k-$cycles $Z_k$ and the sets of $k-$boundaries $B_k$ form abelian subgroups of $C_k$ under addition. It is also to be noted that $ker({\partial_k}) \subset im({\partial_{k+1}})$ since $k-$boundaries are also $k-$cycles. Thus, the groups $B_k$, $Z_k$, and $C_k$ form a nested structure with $B_k \subset Z_k \subset C_k$.

$k$-th homology group $H_k$ is defined as the quotient group, i.e., group of cycles modulo boundaries as given by
\begin{equation}
H_k \;=\; \frac{Z_k}{B_k} \;=\; \frac{ker({\partial_k})}{im({\partial_{k+1}})}\;.
\end{equation}

$H_k(K)$ is the quotient vector space whose generators are given by $k-$cycles that are not boundaries of any $(k\!+\!1)$-simplices. The rank of $H_k(K)$ is also termed as the $k$-th betti number $\beta_k(K)$. $\beta_k(K)$ is the number of $k-$dimensional holes in the simplicial complex $K$ that are not boundaries of any $(k\!+\!1)$-simplices. Here $\beta_0(K)$ is the number of connected components in $K$. It is worth pointing out that betti number $\beta_k$ is an important property of the system that we will make use of in our later analysis. Betti numbers are also used to define Euler characteristics $\chi$, which is a topological invariant of the simplicial complex.
\begin{equation}
\chi \;=\;  \sum_{i=0}^{n}\; (-1)^k \;\beta_k \;.
\end{equation}
\subsection{Persistent Homology and Filtration}
\begin{figure}
    \centering
    \subfloat[]{\includegraphics[width=0.15\textwidth]{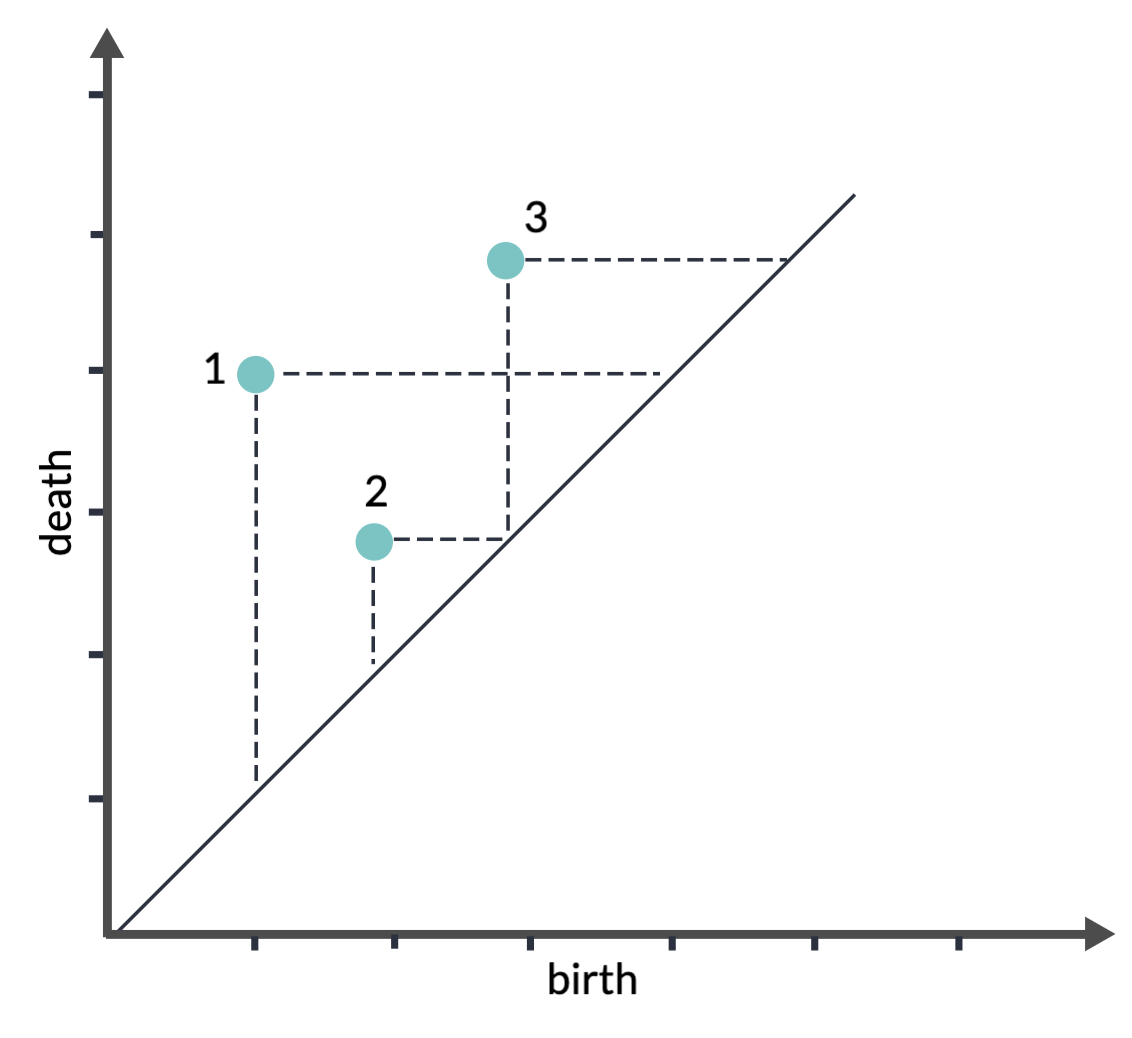}} 
    \subfloat[]{\includegraphics[width=0.15\textwidth]{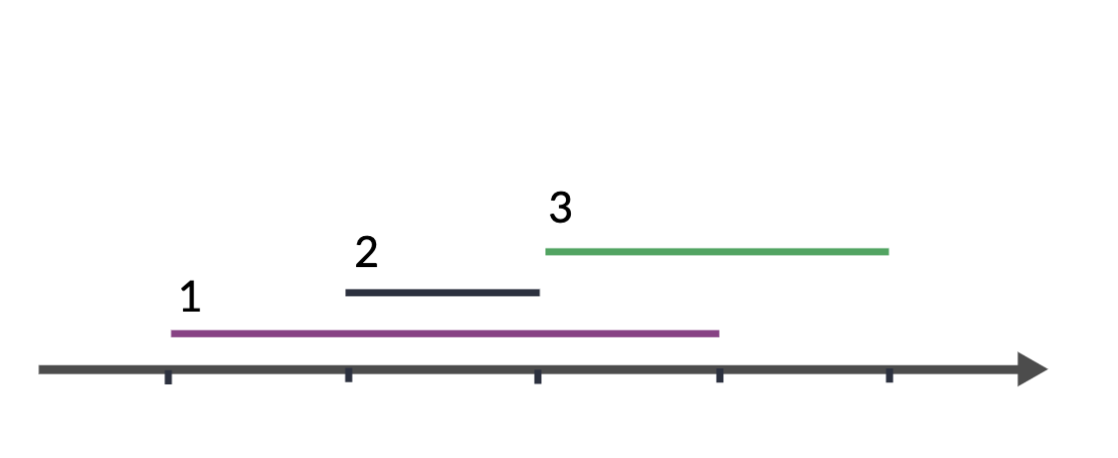}} 
    \subfloat[]{\includegraphics[width=0.15\textwidth]{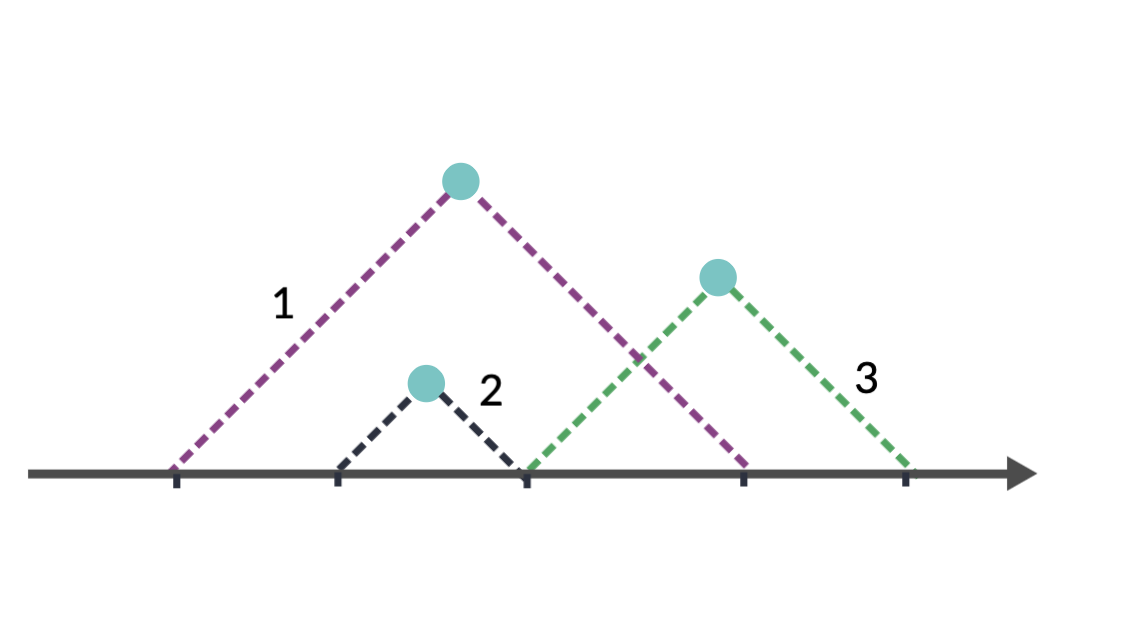}}
    \caption{(a) Persistent Diagram (PD) (b) Barcode Diagram (BD) (c) Persistent Landscape (PL)}
    \label{fig:pd_example}
\end{figure}
The classical homology does not give rich information for a given point cloud data $X$. However, with a multi-scale approach to
homology through filtration parameter,  a change in the homology of $X$ can be captured. The topological features that persist longer are more reliable features of the point cloud data. Instead of working with point cloud data, one considers a family of simplicial complexes $K^\delta$ parameterized by $\delta \in R$ from the set $X$ such that a simplicial complex $K^{\delta_i}$ at step $i$ is a subset of $K^{\delta_j}$ at step $j$ for $i \leq j$. This family of nested simplicial complexes is termed filtration and $\delta$ a filtration parameter that evolves at every $i$-th step. For our case, we will use Vietoris-Rips complex; thus, the natural choice of filtration parameter becomes the radial separation $r$ between points. 

One can keep track of the $birth$-the moment a hole first appears and $death$-the moment a hole disappears in a filtration. Tracking the emergence and disappearance of these homological properties in $K^\delta$ for various values of $\delta$ is the key idea behind persistent homology. These so-called $birth$-$death$ charts are typically represented by \textit{barcode diagram} (BD), \textit{persistent diagram} (PD), \textit{persistent landscape} (PL), \textit{betti curve} (BC), etc.

In figure \ref{fig:pd_example}, we present a simple illustration of three points appearing in the \textit{persistent diagram} shown on the left. For example, point-1 appears at $r=1$ and vanishes at $r=4$, whereas point-2 appears at $r=2$ and vanishes at $r=3$. Thus, point-1 is more persistent than point-2. Thus, the points lying far off the diagonal line on \textit{birth}-\textit{death} chart are more persistent and are true features of the data. The same information can be rendered as one-dimensional \textit{barcode diagram} as shown in the middle image of figure \ref{fig:pd_example}. It can also be represented through \textit{persistent landscape} (figure \ref{fig:pd_example}(c)) when  \textit{persistent diagram} is rotated by $\pi/4$.

Another important measure is the entropy of the points clustered in a \textit{persistence diagram}. It is called \textit{persistence entropy}. Let $D=\{ (b_i,d_i) \}$ be the set of all $birth$-$death$ pairs associated with $k-$th order homology group in \textit{persistence diagram} with $d_i < \infty$. The $k-$th order \textit{persistence entropy} is given by
\begin{equation}
    S(D_k)=-\sum_{i}p_i \log(p_i), 
\end{equation} 
where $p_i=\frac{d_i-b_i}{L_D}$ and $L_D=\sum_{i}{(d_i-b_i)}$.

Similarly, another useful feature associated with persistent homology is \textit{Betti curve}. It is a function $\beta(D_k): R \rightarrow N$, that counts the multiplicity of points of $k-$th homology group at a particular filtration parameter $r=s$ in a \textit{persistence diagram} such that $b_i \leq s \leq d_i$. The area under \textit{Betti curves}, also termed as \textit{Betti area} is a commonly used feature in TDA. 

\begin{figure}[!t]
    \centering
    \subfloat[]{\includegraphics[width=0.15\textwidth]{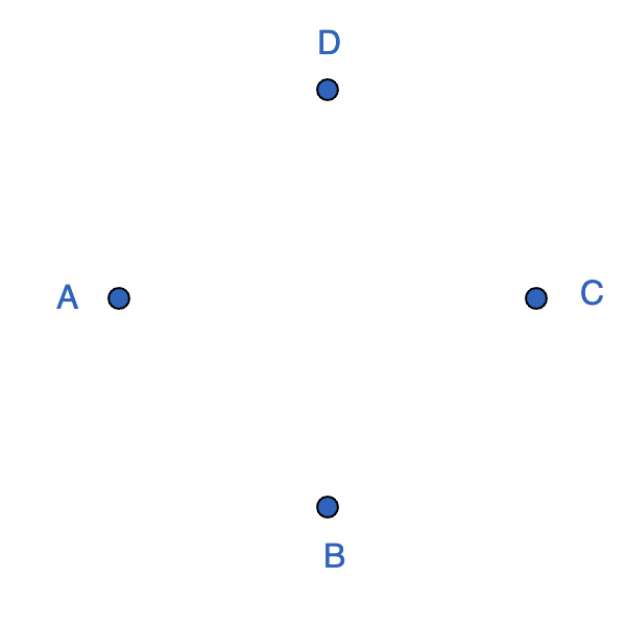}}
    \subfloat[]{\includegraphics[width=0.15\textwidth]{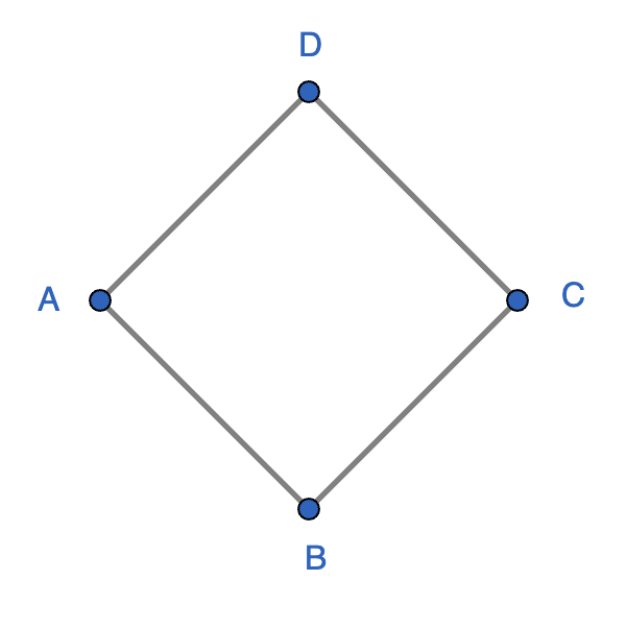}} 
    \subfloat[]{\includegraphics[width=0.15\textwidth]{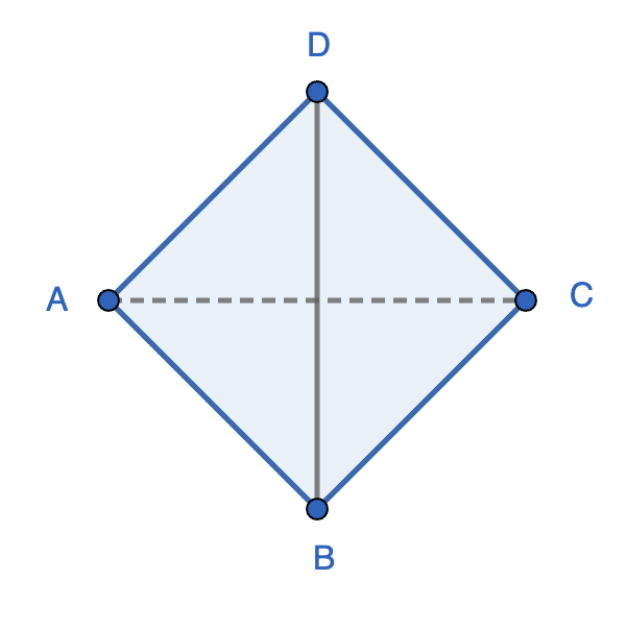}}
    \caption{Rips filtration of point cloud having four points: \{A (-2,2), B (0,0), C (2,2), D (0,4)\} for (a) $r=0$ (b) $r=2\sqrt{2}$ (c) $r=4$.}
    \label{fig:rips_example}
\end{figure}

\subsection{An illustrative example}

In order to illustrate the concepts described above, we take a point cloud consisting of four points: A (0,0), B (0,-2), C (2,0) and D (0,4). We have given the mathematical formulation for forming Rips simplicial complex in equation \ref{eq:rips}. It says that circles of radius $\frac{r}{2}$ centered around each of the $n$ points with a common intersection will form a $(n-1)$-simplex. The Rips-filtration of the example has three important filtration parameters given by $r=0$, $r=2\sqrt{2}$ and $r=4$. 

At $r=0$, the simplicial complex is given by 0-simplices (points) as shown in figure \ref{fig:rips_example}(a). This corresponds to four connected components. The dataset acquires 1-simplices (edges) when $r$ equals the edge length, i.e., $r=2\sqrt{2}$ as depicted in figure \ref{fig:rips_example}(b). At $r\geq 4$, all four points form a clique complex and the simplicial complex is given by the 3-simplex along with all its faces. For $r \geq 2\sqrt{2}$, the number of connected component reduces from four ($\beta_0=4$) to one ($\beta_0=1$).  

For $2\sqrt{2} \leq r < 4$, there is a 1-cycle given by the formal sum of all edges, i.e., $\braket{AB}+\braket{BC}+\braket{CD}+\braket{DA}$. For this choice of $r$, this 1-cycle is not the boundary of any higher order simplex present in the simplicial complex. However, for $r\geq4 $, the 1-cycle $\braket{AB}+\braket{BC}+\braket{CD}+\braket{DA}$ becomes one of the boundaries of the 3-simplex $\braket{ABCD}$. Thus, it does not contribute to the first homology group $H_1$ for $r\geq 4$.
Thus, $\beta_1=1$ for $2\sqrt{2} \leq r < 4$ and $\beta_1=0$ for all other filtration values. The filtration can be summarized as follows.
% For $r < 2\sqrt{2}$ or $r \geq 4$, $\beta_0$ is the only non-zero Betti number and all other Betti numbers are zero. Since $\beta_0$ is the number of connected components

% This is because there is no higher order cycle associated with the simplicial complex. $r < 2\sqrt{2}$, the 1-simplices have not formed and the number of connected components are given by the four vertices. Thus, for $r < 2\sqrt{2}$, $\beta_0=4$. 

% On the otherhand, $r \geq 2\sqrt{2}$, the number of connected components reduces to 1 and this results in $\beta_0=1$. However, for $ 2\sqrt{2} \leq r < 4$, a 1-cycle that is not a boundary appears. Thus, it gives $\beta_1=1$. This is summarized below.

\begin{table}[!ht]
\centering
\begin{tabular}{|l |c|c |c|} 
 \hline 
  Filtration parameter ($r$) & $\beta_0$ & $\beta_1$ & $\chi$ \\ \hline
  $r < 2\sqrt{2}$ &  4  & 0 & 4 \\ \hline
  $ 2\sqrt{2} \leq r < 4 $ &  1  & 1 & 0 \\ \hline
  $r \geq 4 $ &  1  & 0 & 1 \\ \hline
\end{tabular}
\caption{$k-$th Betti numbers ($\beta_k$) and corresponding Euler characteristics $\chi$ for different filtration parameters ($r$) associated with the example.}
\label{tab:betti_example}
\end{table}

As far as $birth$-$death$ persistent diagram is concerned, we have two $birth$-$death$ pairs corresponding to $H_0$ given by $(0,2\sqrt{2})$ with $\beta_0=4$ and $(0,\infty)$ with $\beta_0=1$. On the otherhand, there is only one $birth$-$death$ pair corresponding to $H_1$ given by $(2\sqrt{2},4)$ with $\beta_1=1$. In our subsequent analyses, $birth$-$death$ pairs with $death$ at $r=\infty$ are not included for simplicity.

With this brief introduction to persistent homology, we are ready to dive into the physics discussion. Next, we will talk about the collider simulation and the subsequent persistent homology analysis.
\section{Framework for analysis}
\label{sec:framework}
\subsection{Collider Simulation}
We first choose the resonant production of the electroweak gauge bosons and the Higgs boson. The $Z$ and $W$ bosons are made to decay leptonically and the Higgs boson is made to decay through $b\bar{b}$ channel. Later in the text, we will also investigate the invisible decay modes of the Higgs boson both in the SM and the SSDM.

The event samples are generated at the lowest order (LO) in perturbation theory using \texttt{MadGraph5 aMC@NLO v3.5.1} \cite{Alwall:2014hca, Frederix:2018nkq} with \texttt{nn23lo1} \cite{ball2017parton} patron distribution function. The generated events correspond to $\sqrt{s}=$13 TeV. The generated patron-level events are showered with \texttt{Pythia v8.309} \cite{Bierlich:2022pfr}. In the presence of extra hard partonic jets and the parton shower, the event generator uses the MLM matching technique with the variables \textit{xqcut} and \textit{qcut} set at appropriate values to prevent double counting of events in the simulated samples. We have used an NLO K-factor of 1.2 while estimating the cross sections for all the processes.

The jet-finding package \texttt{FastJet (v3.3.4)} \cite{Cacciari:2005hq,Cacciari:2011ma} included in \texttt{Delphes v3.5.0} \cite{deFavereau:2013fsa} is used to find the jets. The anti-$k_T$ jet algorithm is employed with the cone size set at 0.5, requiring a minimum $p_T^{jet}$ of 20 GeV and the pseudorapidity in the range $|\eta_{jet}| < 2.5$. 

As per the default parameter settings of \texttt{Delphes v3.5.0}, the leptons (electrons and muons) are reconstructed with a minimum $p_T^{l}$ of 0.5 GeV and with $|\eta_{jet}| < 2.5$. The track isolation requirement for electrons and muons involves removing jets that lie within an angular distance $\Delta R \leq 0.5$ from the lepton. Also, to increase the purity of electrons, it is required that the ratio of total $p_T$'s of the stray tracks within the cones of their identification to their own $p_T$'s is less than 0.12. The corresponding ratio for muon is set at 0.25.

% The corresponding requirement for the muons is that the
% maximum total $p_T$ of other tracks does not exceed 1.8 GeV.

Subsequently, the events are processed through fast detector simulation using \texttt{Delphes v3.5.0}. It results in events in \texttt{ROOT} \cite{brun1997root} format and thus, we convert it to \texttt{LHCO} format for further processing. 

\subsection{Point cloud data for TDA}
After the detector level reconstruction step, we require some basic kinematic cuts on the events before we prepare point cloud data from the ensemble of events. From a purely data science perspective guided by collider physics, we extract ($\eta$, $\phi$, $p_T$, $m_{inv}$) from the event files stored in \texttt{LHCO} format as coordinates to represent the entire data as a point cloud. Since these four variables span very different ranges of values, we normalize each of these four variables to lie in the [-1,1] range. It is to be noted here that we are representing the complete ensemble of events along with all the constituent particles that have been reconstructed after fast detector-level simulation. The number of events is normalized to the integrated luminosity times the effective cross-section of the process under consideration. We choose 100 fb$^{-1}$ of integrated luminosity all through the study.

For analysis of the persistent homology of these simulated event bins, we use \texttt{giotto-tda v0.6.0} \citep{giotto-tda}, a high-performance topological machine learning toolbox in \texttt{Python}. The traditional Rips complex described in section \ref{sec:ph} forms a lot more simplices for a given filtration parameter $r$ compared to the Alpha complex. Alpha complex is a Rips complex constructed from the finite cells of a Delaunay triangulation. Thus, we choose the latter in order to reduce the computational cost. This corresponds to \texttt{WeakAlphaPersistence} module in \texttt{giotto-tda} package.

% We use Vietoris-Rips complex algorithm described previously to build the simplicial complexes and filtration from the data points represented by normalized ($\eta$, $\phi$, $p_T$, $m_{inv}$) coordinates. This results in \textit{birth}-\textit{death} chart or persistent diagrams for each bin. We determine the average \textit{persistent diagram}, average persistent entropy and average Betti areas for up to the second homology dimension.

\section{Persistent homology of the SM}
\label{sec:ph-sm}
With the above-mentioned framework of analysis, we now delve into the study of the persistent homology of the electroweak sector of the SM. We consider resonant production of \textit{ZZ}, \textit{ZW$^{\pm}$},  \textit{$W^+W^-$}, \textit{ZH} and \textit{W$^{\pm}$H} processes in the SM.
We apply $p_T^{l}>[50,40,10,10]$ GeV depending on the lepton counts. For the SM processes involving $b\bar{b}$ decays, we keep $p_T^{b}> 30$ GeV. For $Z\rightarrow l^+ l^-$ processes, we keep the leptonic invariant mass window from 80 GeV to 100 GeV. For ZW and WH processes, we keep leptonic transverse mass $m_T^{l_1} > 80$ GeV and for WW, stransverse mass $m_{T2}^{l_1,l_2}> 70$ GeV.

% \begin{figure}
% \centering
%   \begin{subfigure}[b]{0.4\textwidth}
%     \includegraphics[width=\textwidth]{bd_sm_100_wh_lvbb.pdf}
%   \end{subfigure}
%  \hfill
%   \begin{subfigure}[b]{0.4\textwidth}
%     \includegraphics[width=\textwidth]{bd_sm_100_zh_llbb.pdf}
%   \end{subfigure}
%     \caption{$birth$-$death$ chart or persistent diagrams for WH (a), ZH (b), WW (c), ZW (d) and ZZ (e) productions in the SM. The legends $H_k$ stand for the $k-$th homology group, i.e., $k-$dimensional holes.}
%     \label{fig:pd_sm}
% \end{figure}

\begin{figure*}[t]
  \centering
  \subfloat[]{\includegraphics[width=0.3\textwidth]{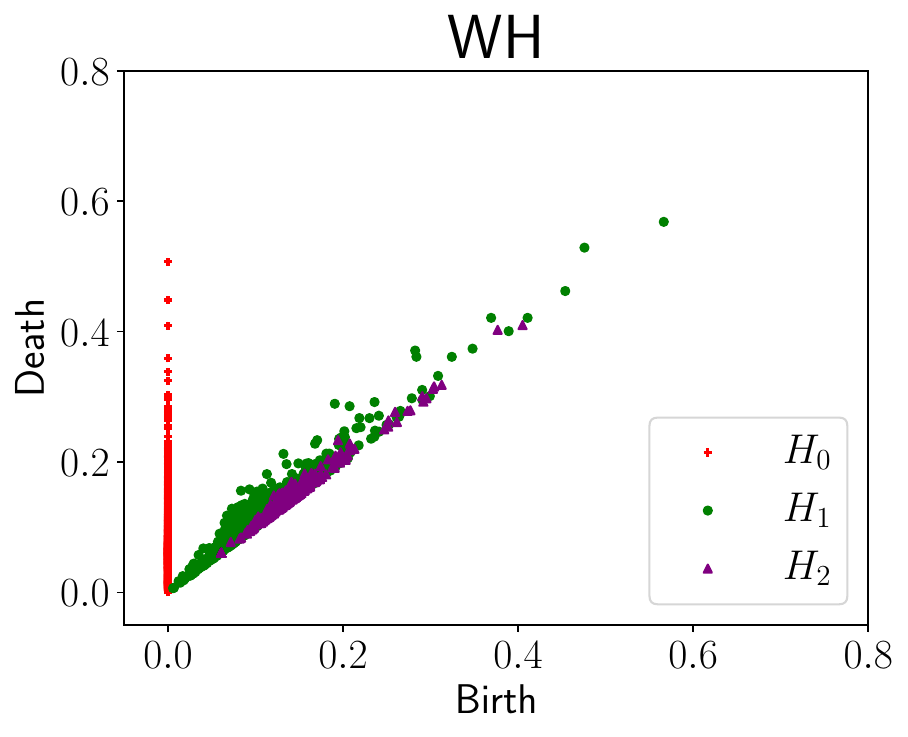}}
  % \hfill
  \subfloat[]{\includegraphics[width=0.3\textwidth]{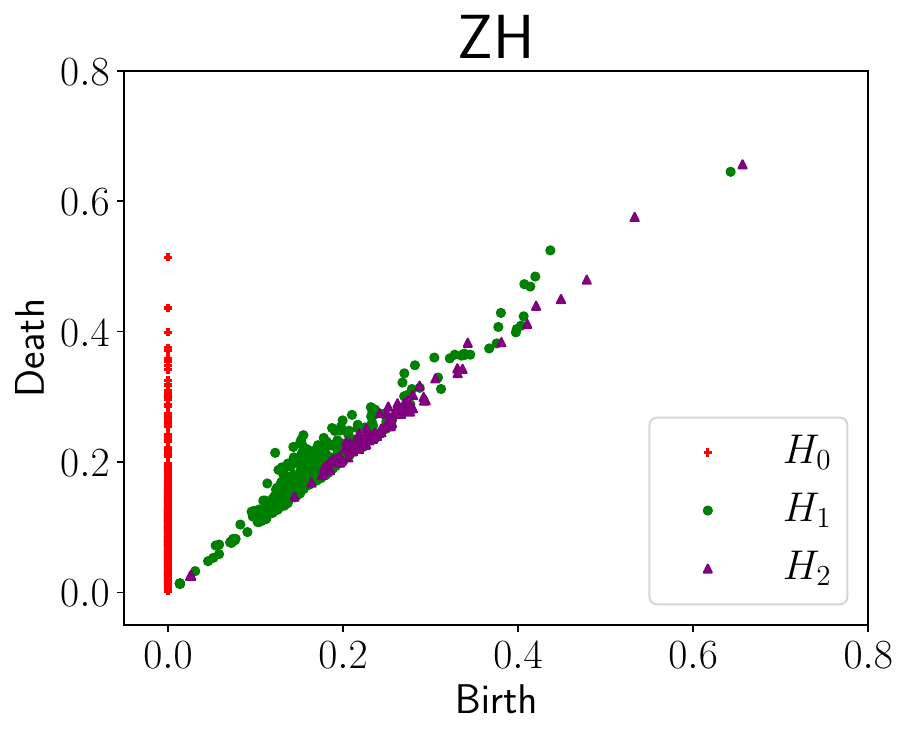}}
  \subfloat[]{\includegraphics[width=0.3\textwidth]{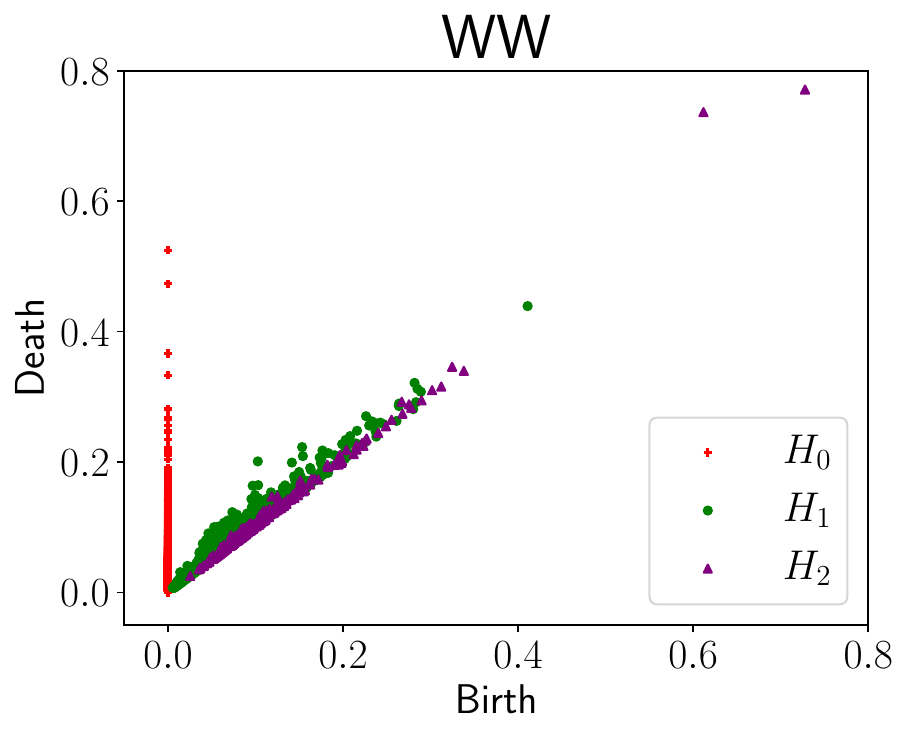}}
  \vspace{1pt} % Adjust the vertical space between rows of subfigures
  \subfloat[]{\includegraphics[width=0.3\textwidth]{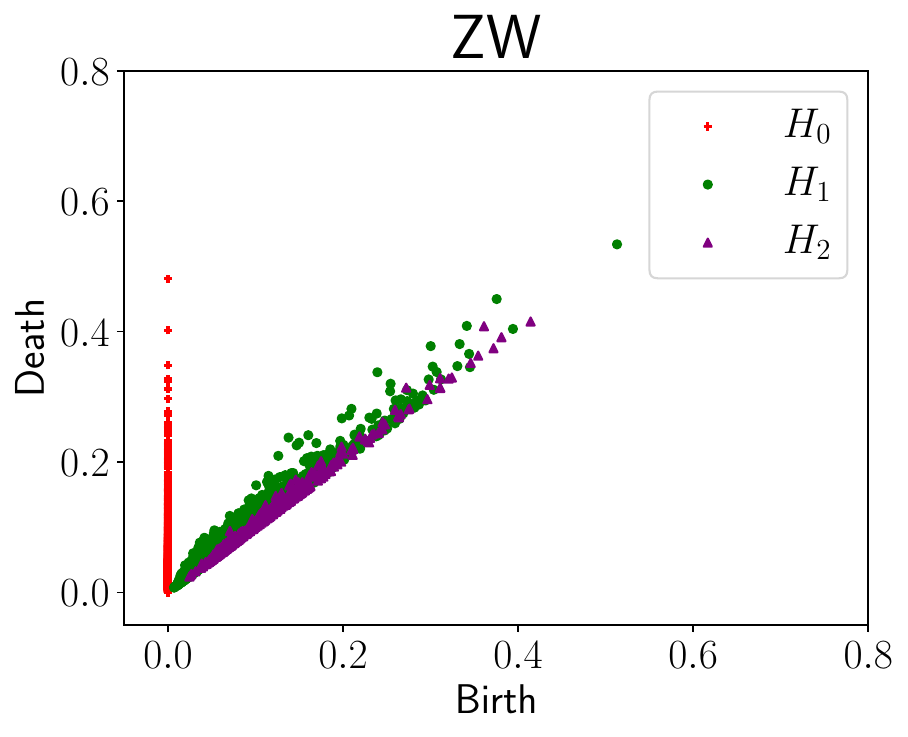}}
  % \hfill
  \subfloat[]{\includegraphics[width=0.3\textwidth]{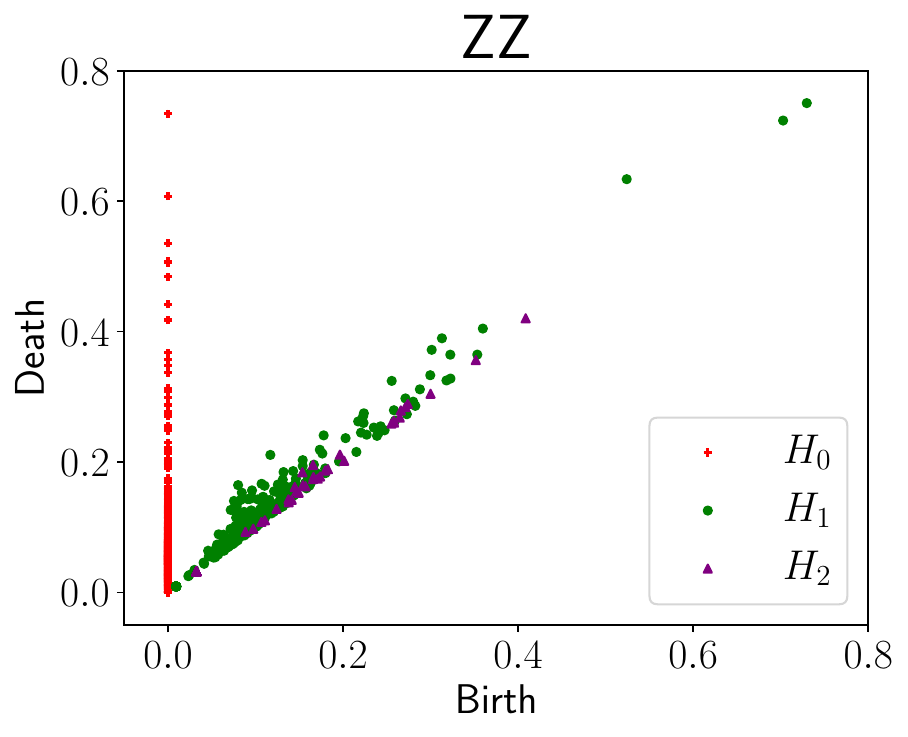}}

  \caption{$birth$-$death$ chart or persistent diagrams for WH (a), ZH (b), WW (c), ZW (d) and ZZ (e) productions in the SM. The legends $H_k$ stand for the $k-$th homology group, i.e., $k-$dimensional holes.}
  \label{fig:pd_sm}
\end{figure*}

% \begin{figure}[!ht]
%     \centering
%     \subfigure[]{\includegraphics[width=0.15\textwidth]{bd_sm_100_wh_lvbb.pdf}} 
%     \subfigure[]{\includegraphics[width=0.15\textwidth]{bd_sm_100_zh_llbb.pdf}} 
%     \subfigure[]{\includegraphics[width=0.15\textwidth]{bd_sm_100_ww_lv.pdf}}
%     \subfigure[]{\includegraphics[width=0.15\textwidth]{bd_sm_100_zw_lv.pdf}} 
%     \subfigure[]{\includegraphics[width=0.15\textwidth]{bd_sm_100_zz_ll.pdf}} 
%     \caption{$birth$-$death$ chart or persistent diagrams for WH (a), ZH (b), WW (c), ZW (d) and ZZ (e) productions in the SM. The legends $H_k$ stand for the $k-$th homology group, i.e., $k-$dimensional holes.}
%     \label{fig:pd_sm}
% \end{figure}

\begin{table}[!ht]
\centering
\begin{tabular}{|l |c|c |c|} 
 \hline 
  Event type & $S_0$ & $S_1$ & $S_2$  \\ \hline
  WH &  1.52  & 2.39 & 10.13 \\ \hline
  ZH & 1.52 &  2.86 & 67.22  \\ \hline
  WW & 1.60 &  2.50 & 10.28 \\ \hline
  ZW & 1.57 & 2.32 & 6.35 \\ \hline
  ZZ & 1.59 & 3.36 & -2.19 \\ \hline
\end{tabular}
\caption{Persistent entropies for the SM resonant productions. $S_k$ stands for $k-$th persistent entropy.}
\label{tab:pe_sm}
\end{table}

\begin{figure*}[!ht]
    \centering
    \subfloat[]{\includegraphics[width=0.3\textwidth]{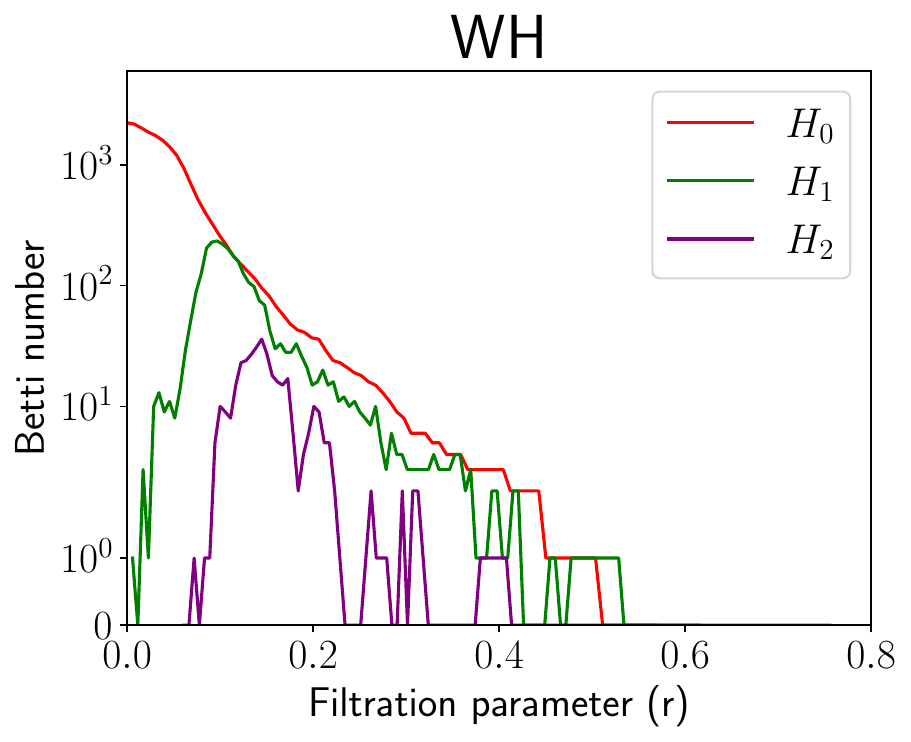}} 
    \subfloat[]{\includegraphics[width=0.3\textwidth ]{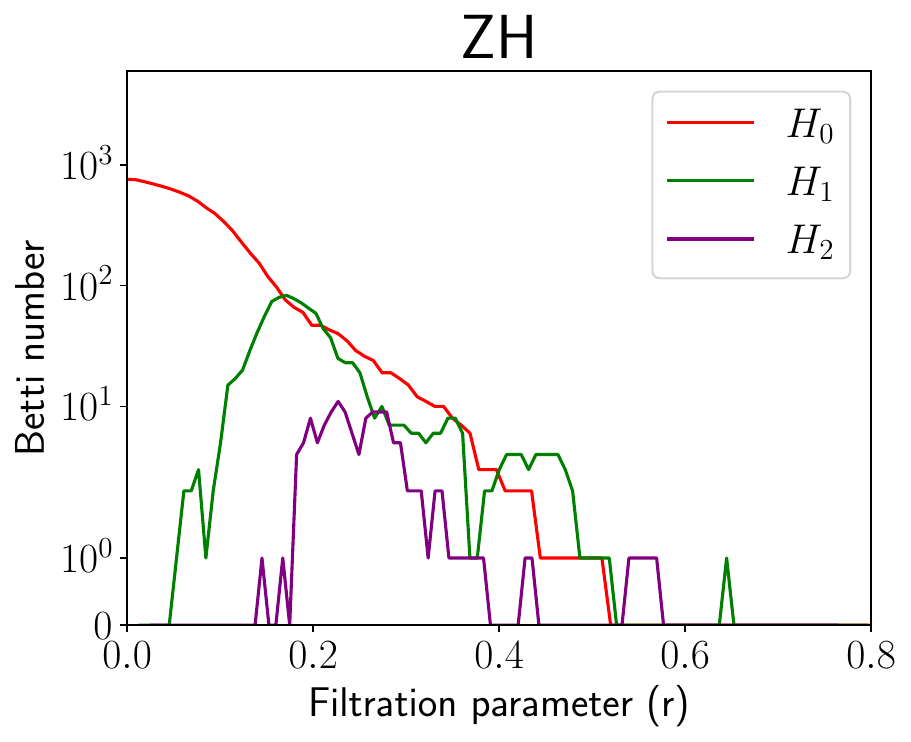}} 
    \subfloat[]{\includegraphics[width=0.3\textwidth]{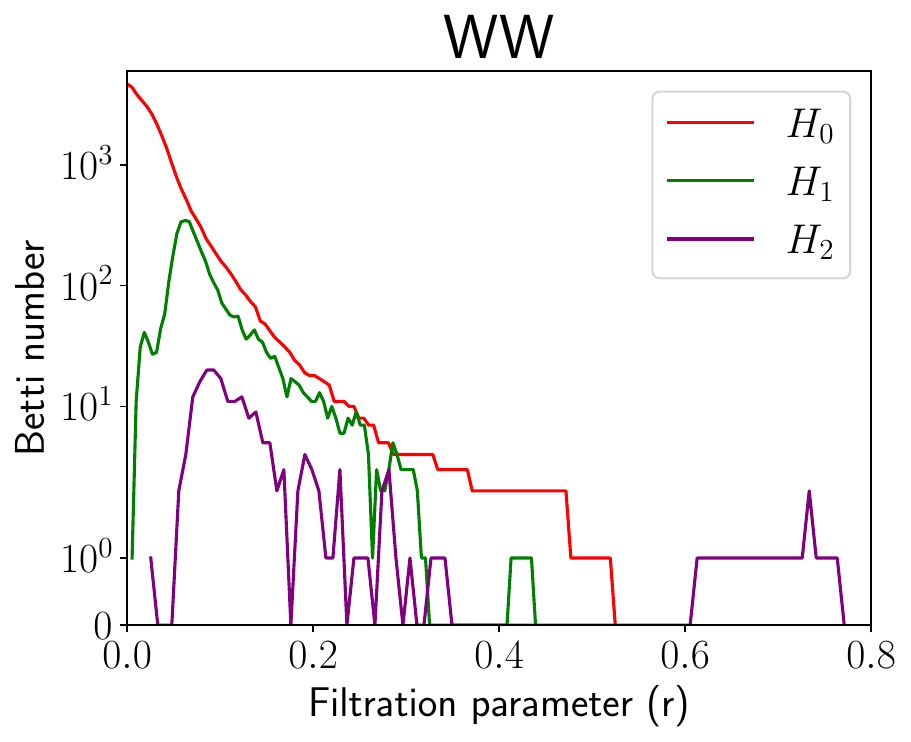}}
    \vspace{1pt}
    \subfloat[]{\includegraphics[width=0.3\textwidth]{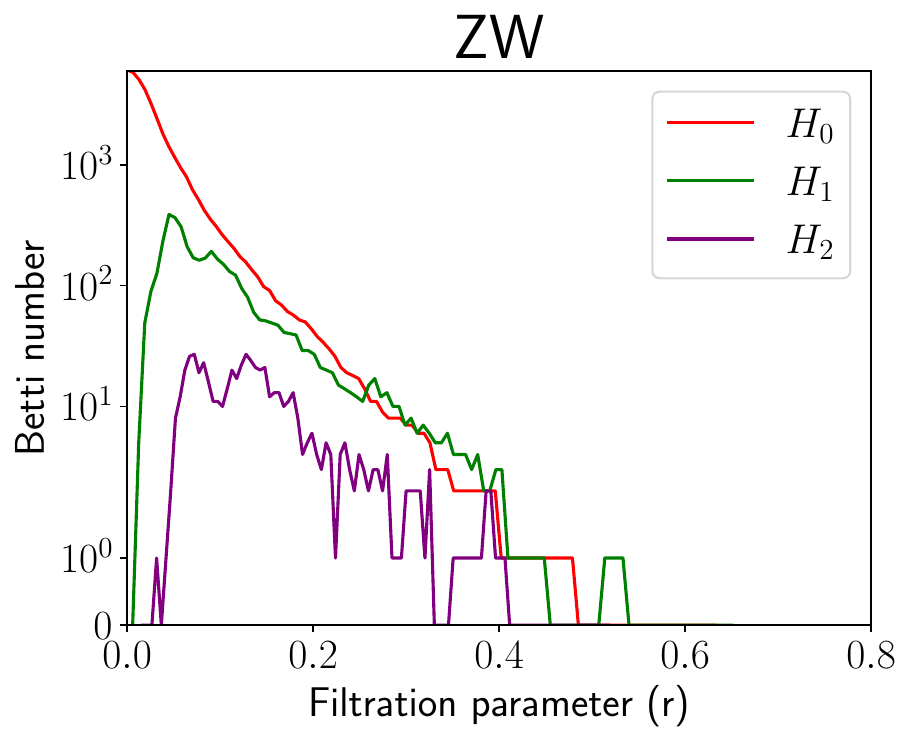}} 
   \subfloat[]{\includegraphics[width=0.3\textwidth]{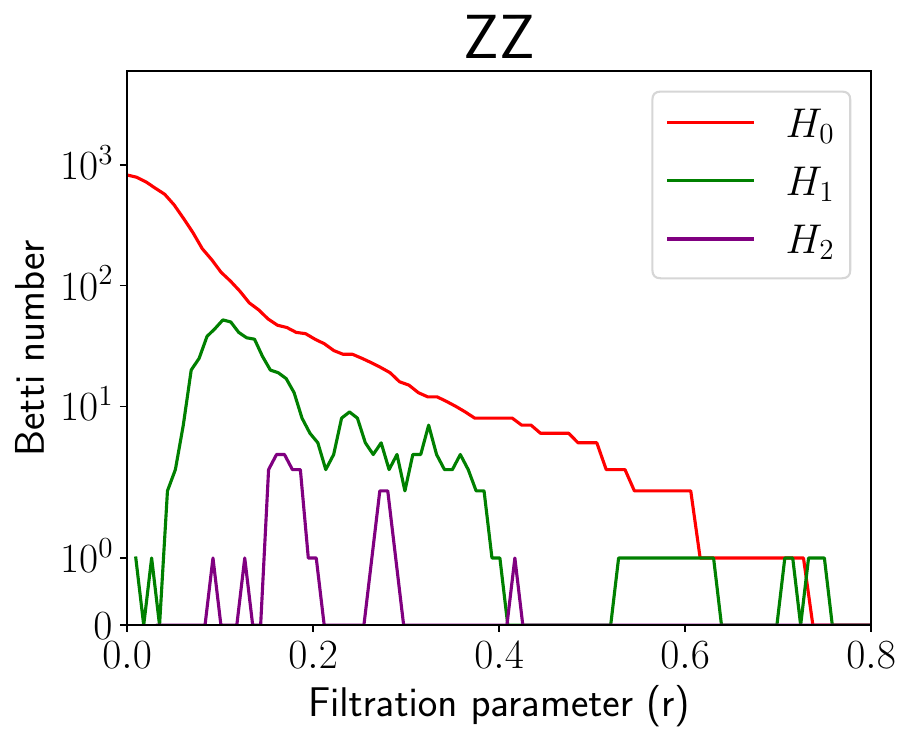}} 

    \caption{Betti curves for WH (a), ZH (b), WW (c), ZW (d) and ZZ (e) productions in the SM.}
    \label{fig:ba_sm}
\end{figure*}

\begin{figure*}[!htp]
    \centering
    \subfloat[]{\includegraphics[width=0.23\textwidth]{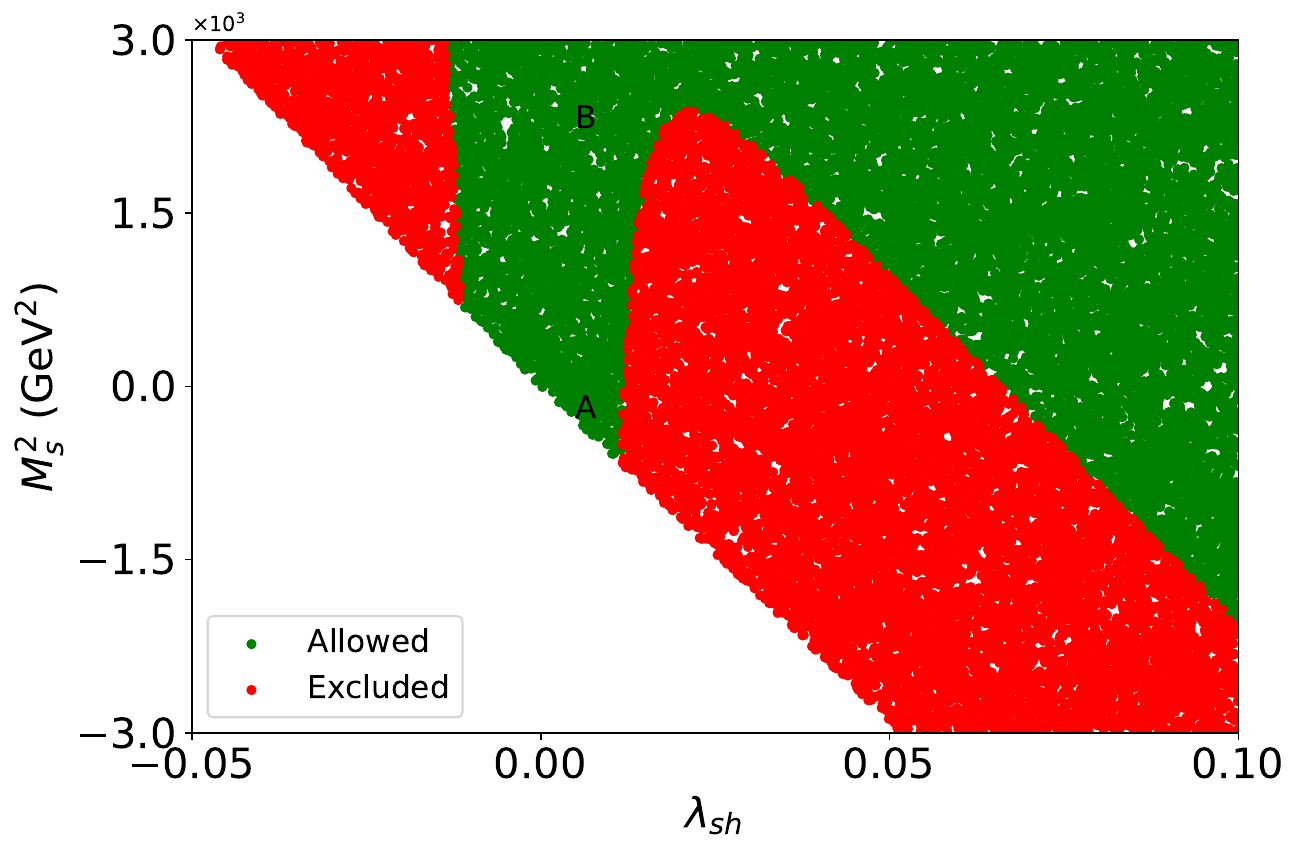}} 
    \subfloat[]{\includegraphics[width=0.23\textwidth]{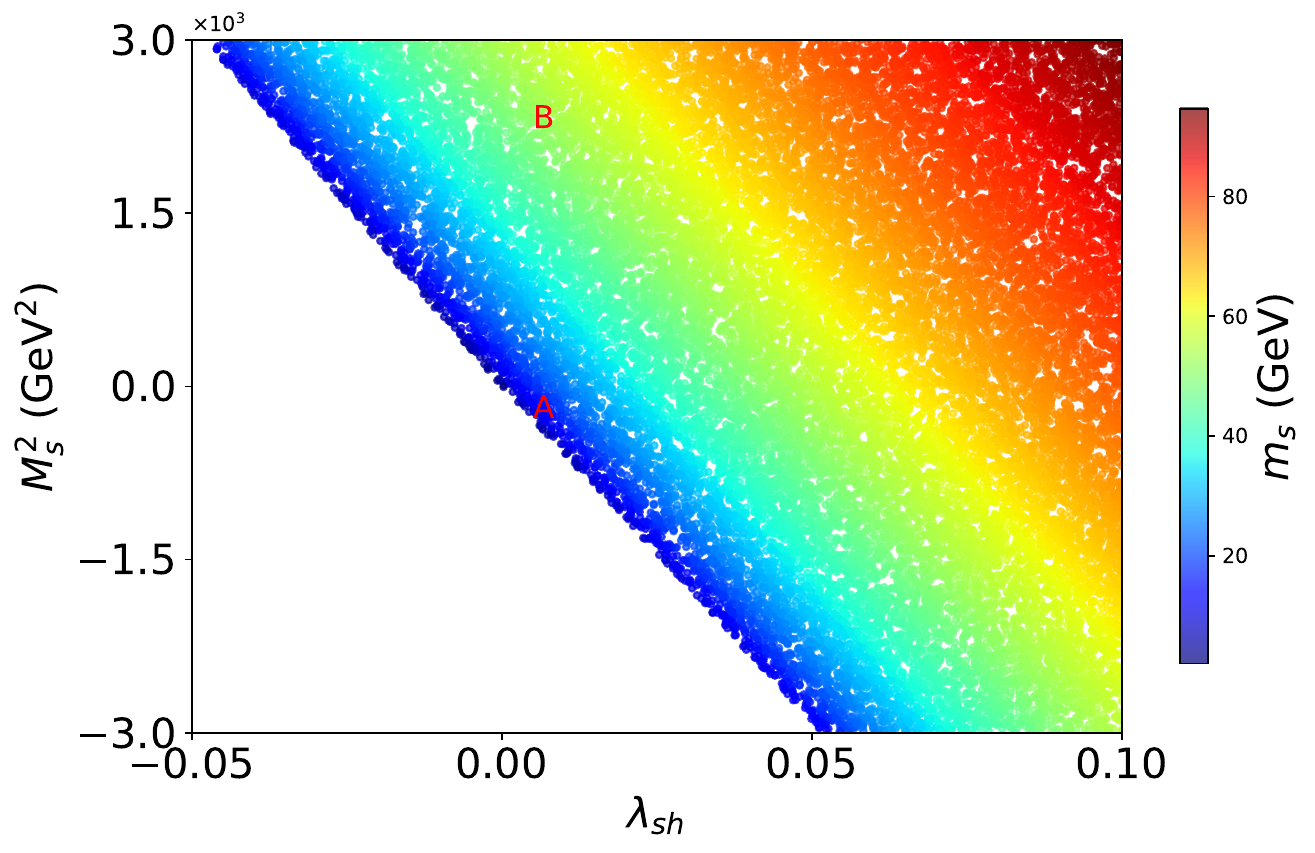}} 
    \vspace{1pt}
    \subfloat[]{\includegraphics[width=0.23\textwidth]{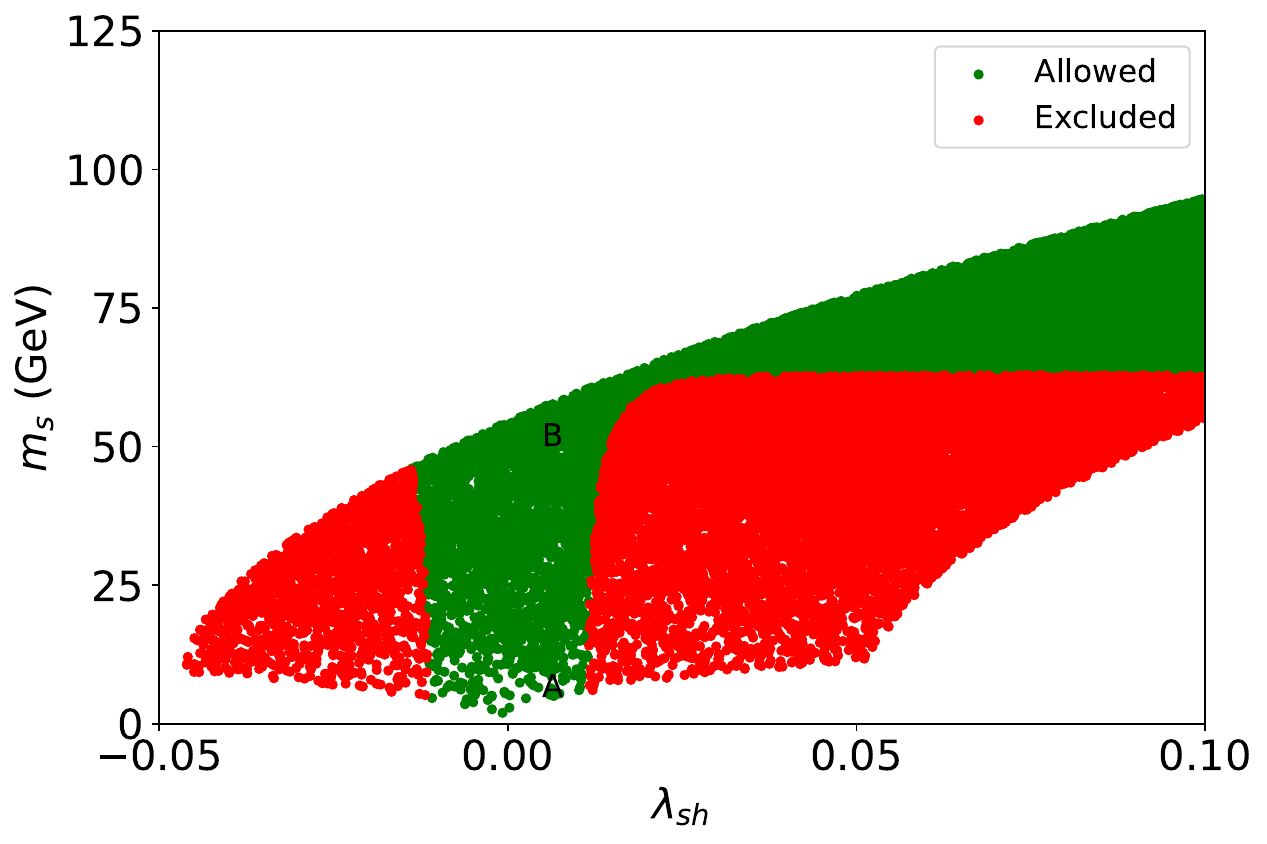}} 
    \subfloat[]{\includegraphics[width=0.23\textwidth]{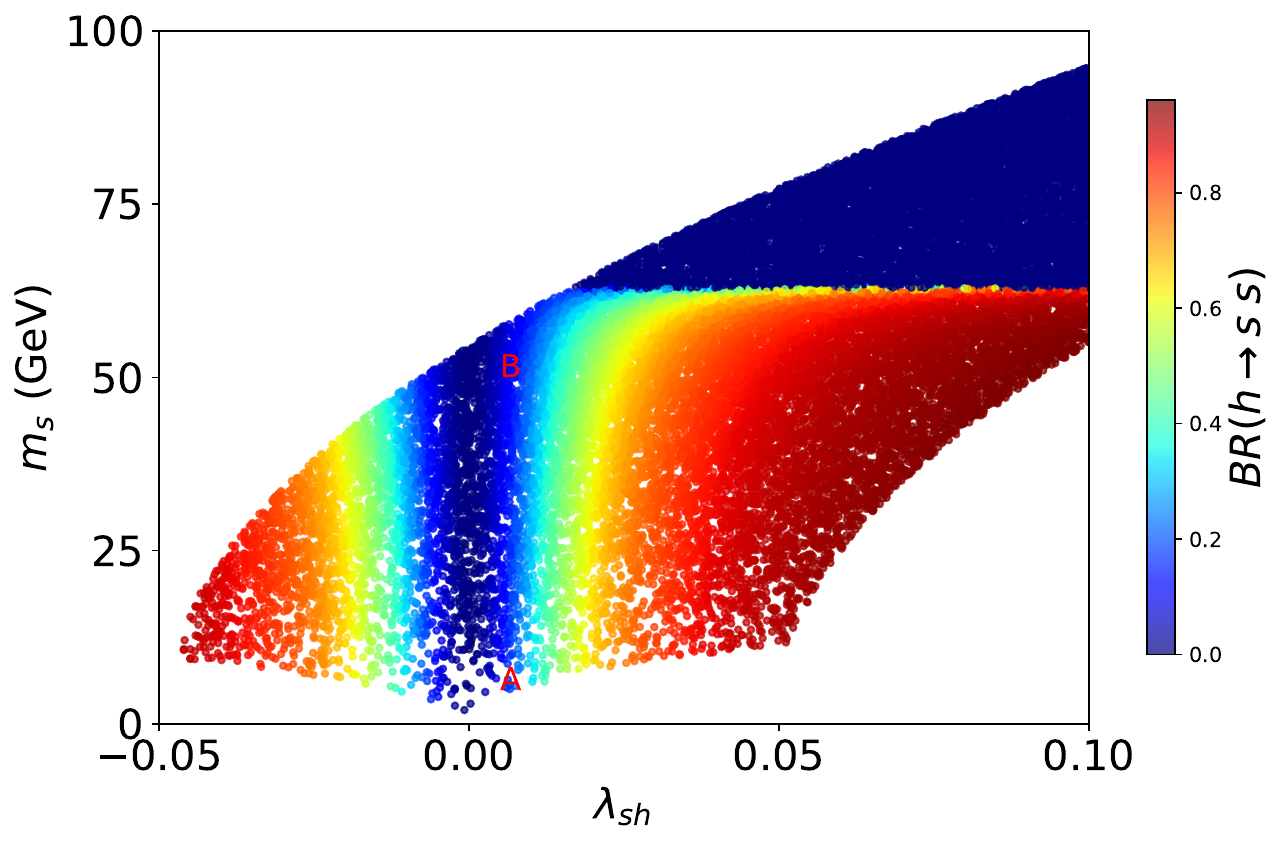}} 
    \caption{Scan over parameter space of the SSDM. The red (green) patches in (a) and (c) refer to excluded (allowed) regions when subjected to constraints from all searches at LHC for a 125 GeV neutral scalar. The $\lambda_{sh}$-$M_S^2$ parameter space with heat map with $m_s$ is shown in (b). The same region with Higgs invisible branching fraction is given in (d). The benchmark points A and B in table \ref{tab:ssdm_benchmarks} are also indicated in all of these plots.}
    \label{fig:scan_ssdm}
\end{figure*}

\begin{table}[!ht]
\centering
\begin{tabular}{|l |c|c |c|c|} 
 \hline 
  Event type & $BA_0$ & $BA_1$ & $BA_2$ \\ \hline
  WH &  135.82 (0.89) &  15.64 (0.10) & 1.98 (0.01)\\ \hline
  ZH & 79.19 (0.90) & 8.12 (0.09) & 1.03 (0.01)\\ \hline
  WW & 164.72 (0.90)  & 17.27 (0.09) & 1.57 (0.01)\\ \hline
  ZW & 219.44 (0.89) & 25.81 (0.10) & 2.78 (0.01)\\ \hline
  ZZ & 58.13 (0.92) & 4.95 (0.08) &	0.29 (0.0)\\ \hline
\end{tabular}
\caption{$BA_k$ is the $k-$th order Betti area for WH, ZH, WW, ZW and ZZ productions in the SM. The fractional Betti areas are indicated in brackets.}
\label{tab:ba_sm}
\end{table}

In figure \ref{fig:pd_sm}, we present persistent diagrams for WH, ZH, WW, ZW, and ZZ corresponding to $H_{0,1,2}$ homology groups. Zeroth, first and second order persistent entropy are represented by red (+), green (dot) and purple (triangle) points. In table \ref{tab:pe_sm}, zeroth order persistent entropy for WH and ZH are the same ($S_0=1.52$) and for WW, ZW and ZZ, $S_0 \approx 1.60$. In figure \ref{fig:pd_sm}(a) and \ref{fig:pd_sm}(b), the purple triangles (corresponds to $H_2$) has $birth$ points beginning at larger $r\approx0.1$ and $r\approx0.2$, respectively compared to figure \ref{fig:pd_sm}(c) and \ref{fig:pd_sm}(d). In table \ref{tab:pe_sm}, we see that $S_1$ for ZZ and ZH are the largest compared to others. Similarly, $S_2$ for ZH is also the largest and it is corroborated with the spread of purple triangles to large $r\approx0.6$.

In figure \ref{fig:ba_sm}, we present Betti curves for WH, ZH, WW, ZW and ZZ corresponding to $H_{0,1,2}$ homology groups. We also present the Betti areas (area under Betti curves) in table \ref{tab:ba_sm} along with the fractional Betti area in the brackets. We see that ZH and ZZ have the lowest Betti areas across all homology dimensions. This is because of lesser number of points in the ensemble resulting from lower production cross-sections compared to others. However, the fractional contributions of Betti areas are somewhat similar in the SM except for a slightly larger (0.92) for the ZZ production. Also, the shape of Betti curves corresponding to $H_{0,1}$ for ZH and ZZ are different from WH, WW and ZW in terms of the location of peak and slope of the curve in figure \ref{fig:ba_sm}. 

Next, we consider the resonant production of WH and ZH and subsequent leptonic decay of gauge bosons and invisible decay of the Higgs boson in a real singlet extension of the SM featuring a scalar dark matter (DM) candidate.

\section{Real singlet extension of the SM}
\label{sec:ssdm}

We consider a simple extension of the SM with a real $SU(2)_{\mathrm{L}}$ singlet scalar field $S$. We also impose $Z_2$ symmetry on the scalar sector such that $S$ is odd and all SM fields are even under that. Since $Z_2$ symmetry ensures a stable singlet scalar particle, this model also serves as the simplest extension of the SM featuring a scalar dark matter (DM) candidate. We choose this model to illustrate that the persistent homology of the ensemble of collider signals of a BSM model can have a different topological signature compared to the SM. The tree-level scalar potential is given by
\begin{equation}
    V(H,S)=-\frac{\mu^2}{2} H^\dagger H - \frac{\lambda_H}{2} (H^\dagger H)^2 
    -\frac{M_S^2}{2} S^2 - \frac{\lambda_s}{2} S^4 - \lambda_{sh} H^\dagger H S^2
\end{equation}
Around the vacuum expectation value ($vev$), the neutral component of the SM Higgs doublet ($\mathcal{H}_0$) and real singlet scalar ($S$) is parameterized as
\begin{equation}
    \mathcal{H}_0=\frac{v+h}{\sqrt{2}},\; S=\frac{v_s+s}{\sqrt{2}},
\end{equation}
where $v=246$ GeV ($v_s$) is the $vev$ for $\mathcal{H}_0$ ($S$). $Z_2$ invariance of the scalar sector requires zero $vev$ for the real singlet scalar, i.e., $v_s=\langle S \rangle=0$. Thus, the tree-level mass of the real singlet $S$ is given by
\begin{equation}
m_s^2=M_S^2 + \lambda_{sh} v^2 
\label{eq:ms_relation}
\end{equation}
Thus, in this simple extension of the SM, the mass of the singlet dark matter (DM) candidate $S$ is primarily governed by $M_S^2$ and  $\lambda_{sh}$. $\lambda_{sh}$ is the only term in the scalar potential contributing to the coupling between singlet DM and the SM Higgs boson. Thus, the branching ratio (BR) of the Higgs for invisible decay to singlet scalar DM is primarily determined by the $\lambda_{sh}$ and $m_h-m_s$.

\begin{figure*}[!htp]
    \centering
    \subfloat[]{\includegraphics[width=0.3\textwidth]{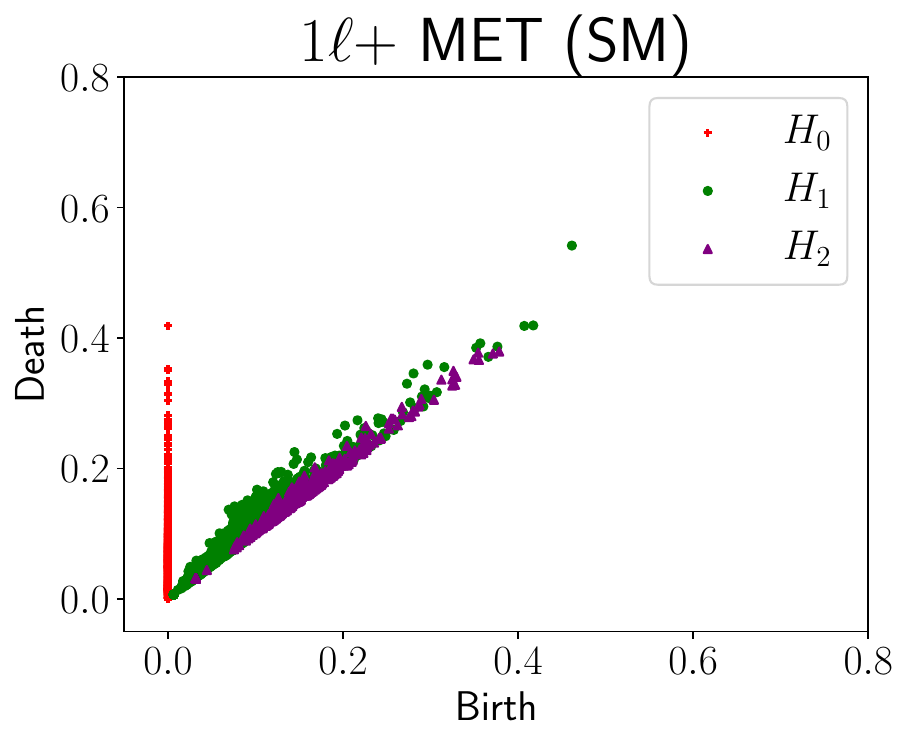}} 
    \subfloat[]{\includegraphics[width=0.3\textwidth]{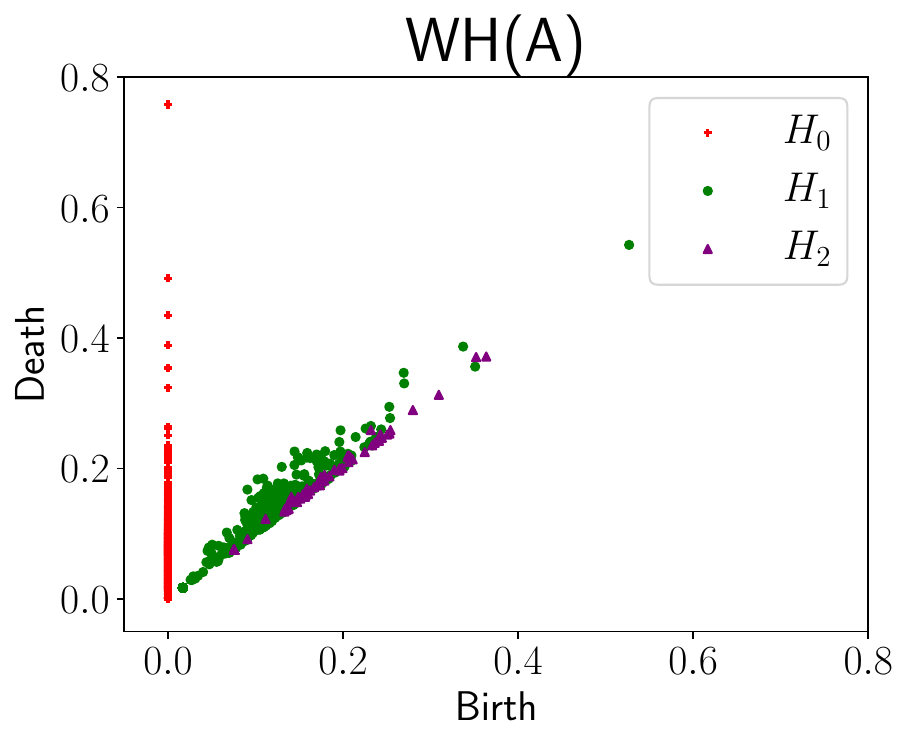}} 
    \subfloat[]{\includegraphics[width=0.3\textwidth]{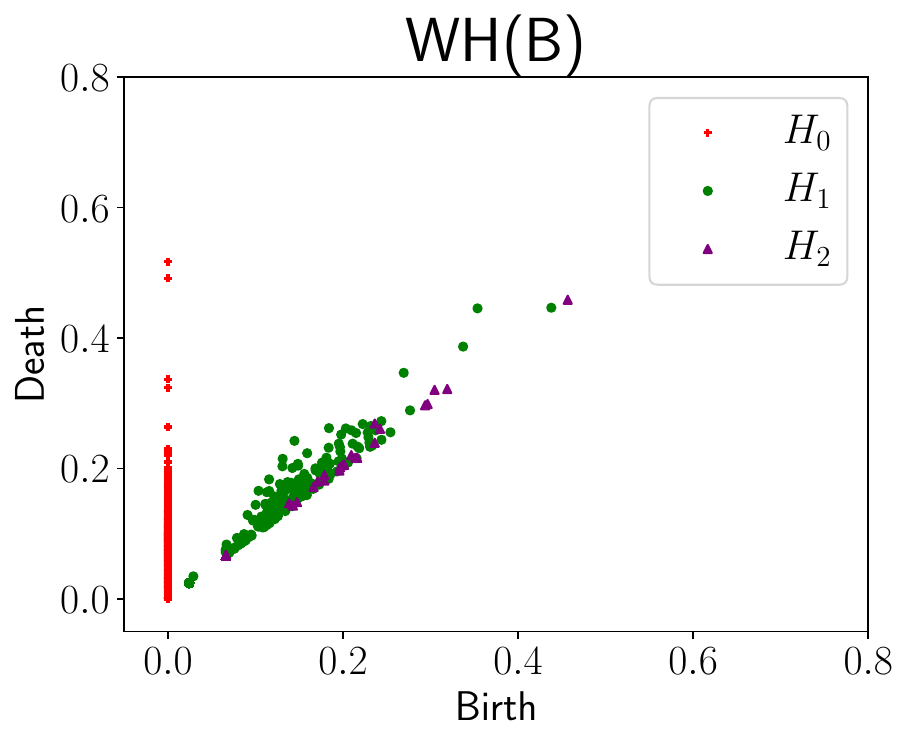}} 
     \vspace{1pt}
    \subfloat[]{\includegraphics[width=0.3\textwidth]{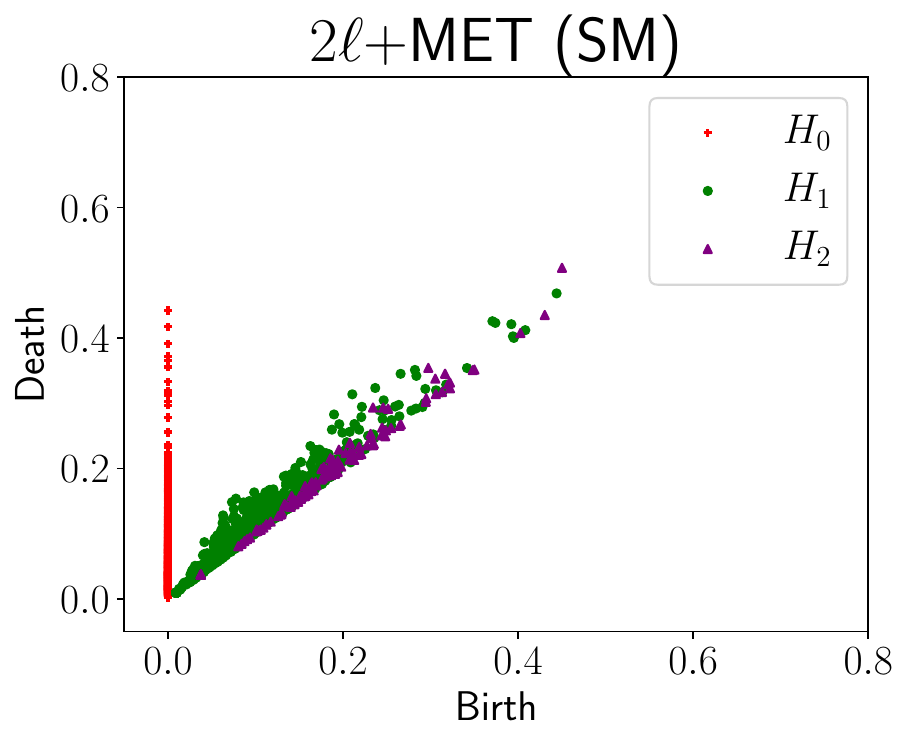}} 
    \subfloat[]{\includegraphics[width=0.3\textwidth]{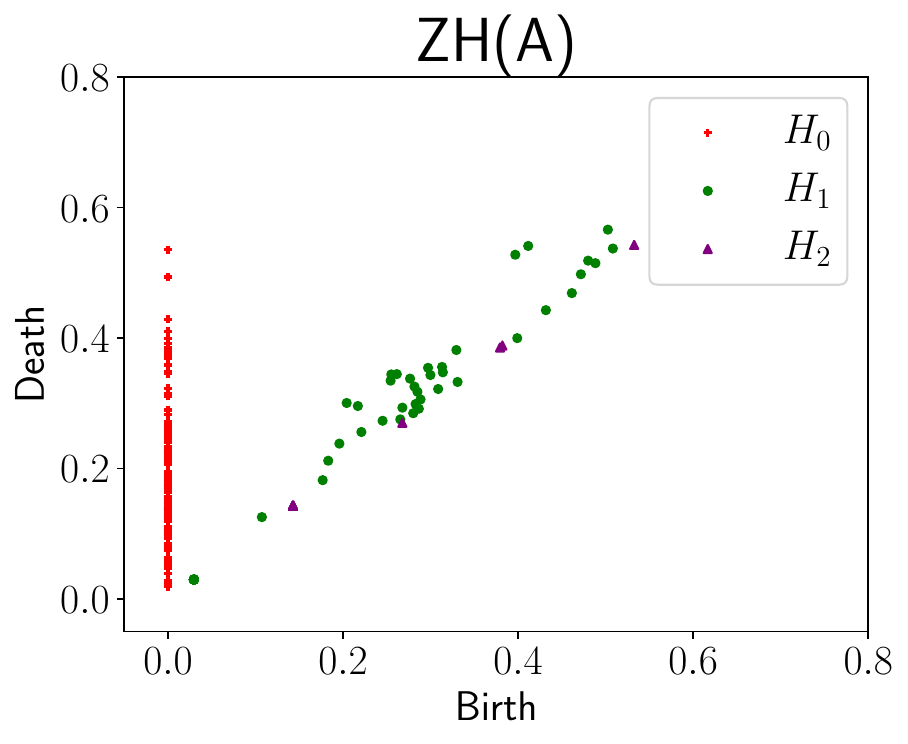}} 
    \subfloat[]{\includegraphics[width=0.3\textwidth]{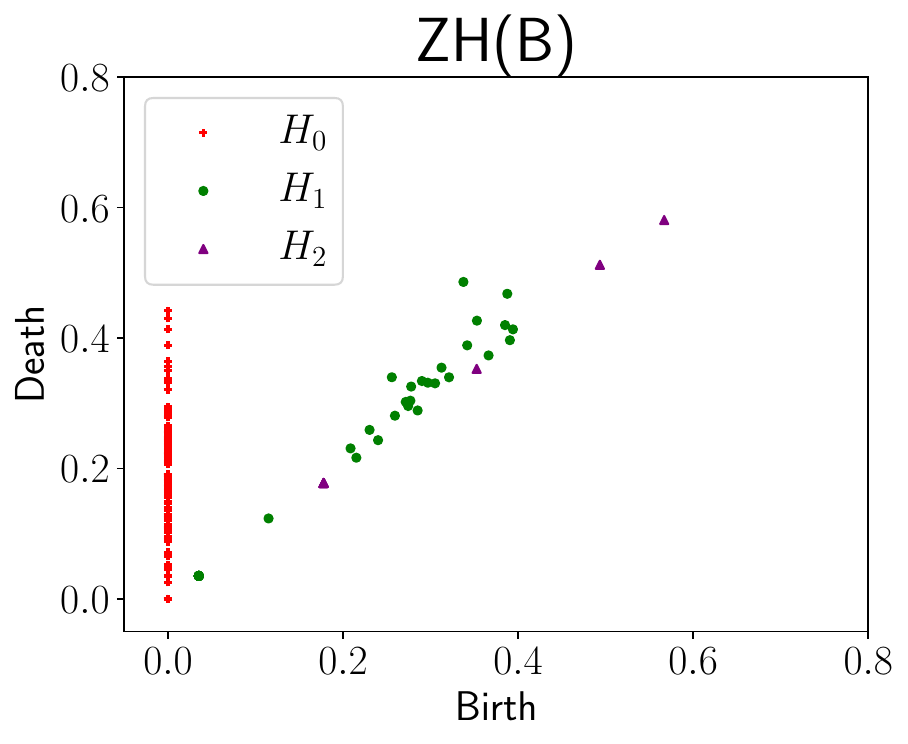}} 
    \caption{$birth$-$death$ charts or the persistent diagrams for the Higgs invisible decay in the SSDM benchmark scenarios and the SM backgrounds are presented. First (second) row corresponds to WH (ZH) production mode in the SSDM.}
    \label{fig:pe_ssdm}
\end{figure*}

\begin{figure*}[!htp]
    \centering
    \subfloat[]{\includegraphics[width=0.3\textwidth]{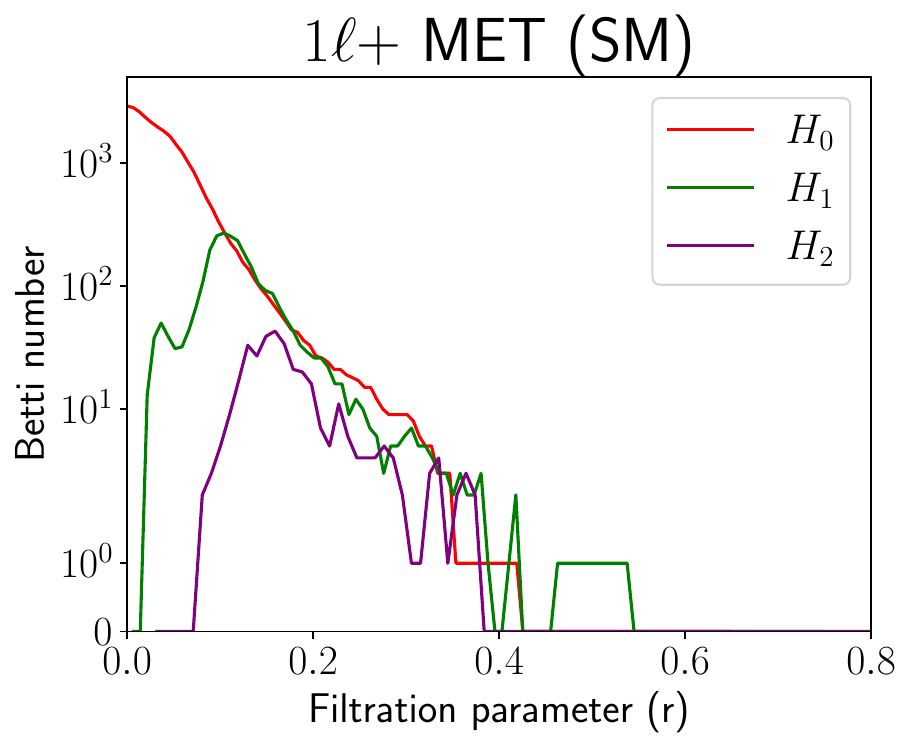}} 
    \subfloat[]{\includegraphics[width=0.3\textwidth]{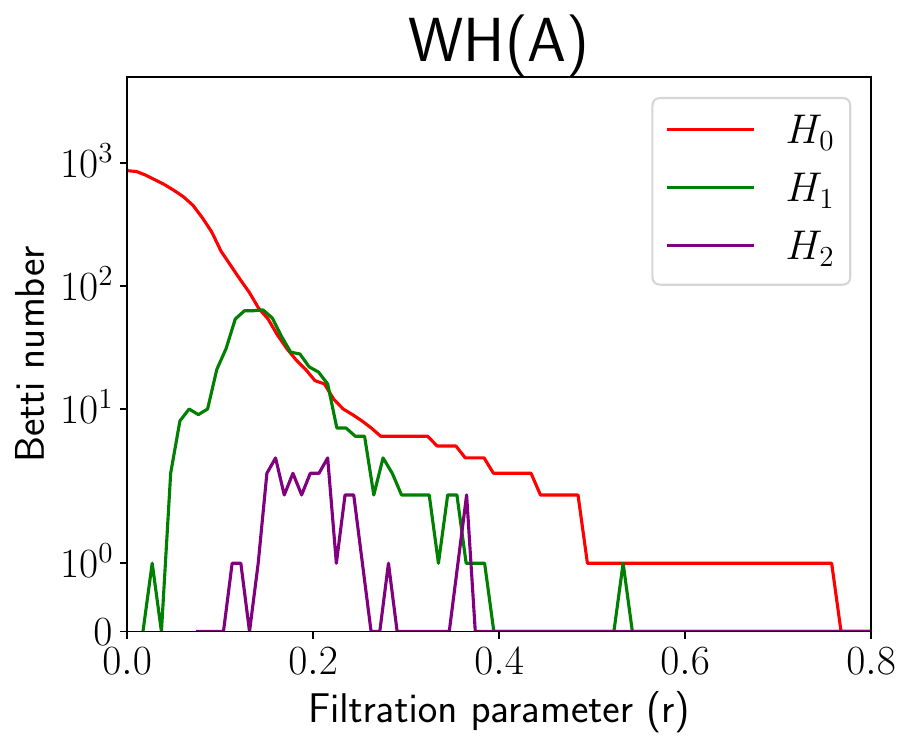}} 
    \subfloat[]{\includegraphics[width=0.3\textwidth]{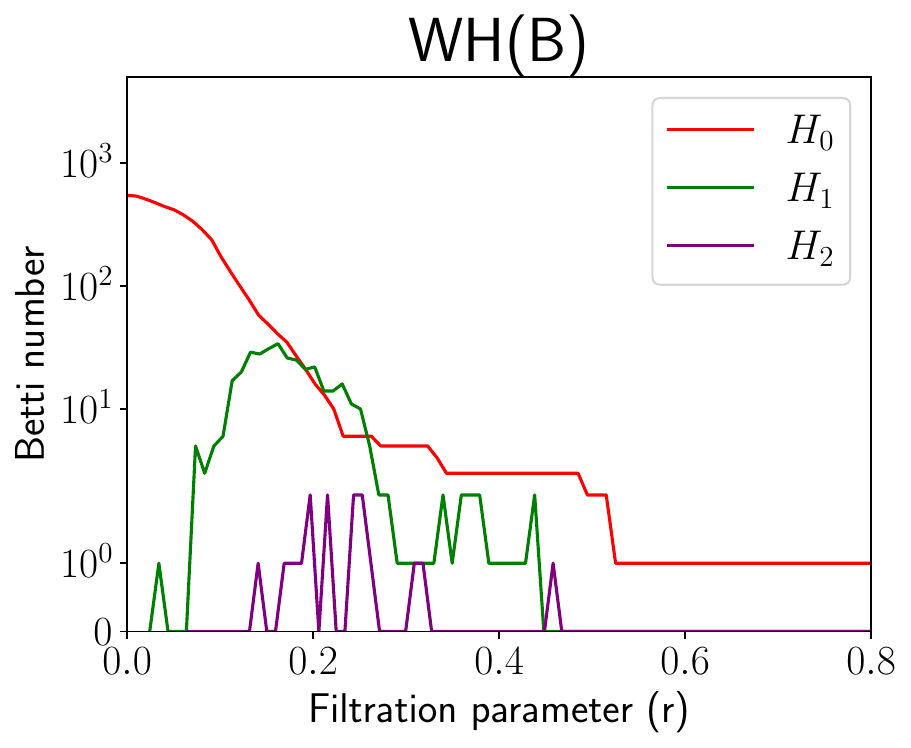}} 
    \vspace{1pt}
    \subfloat[]{\includegraphics[width=0.3\textwidth]{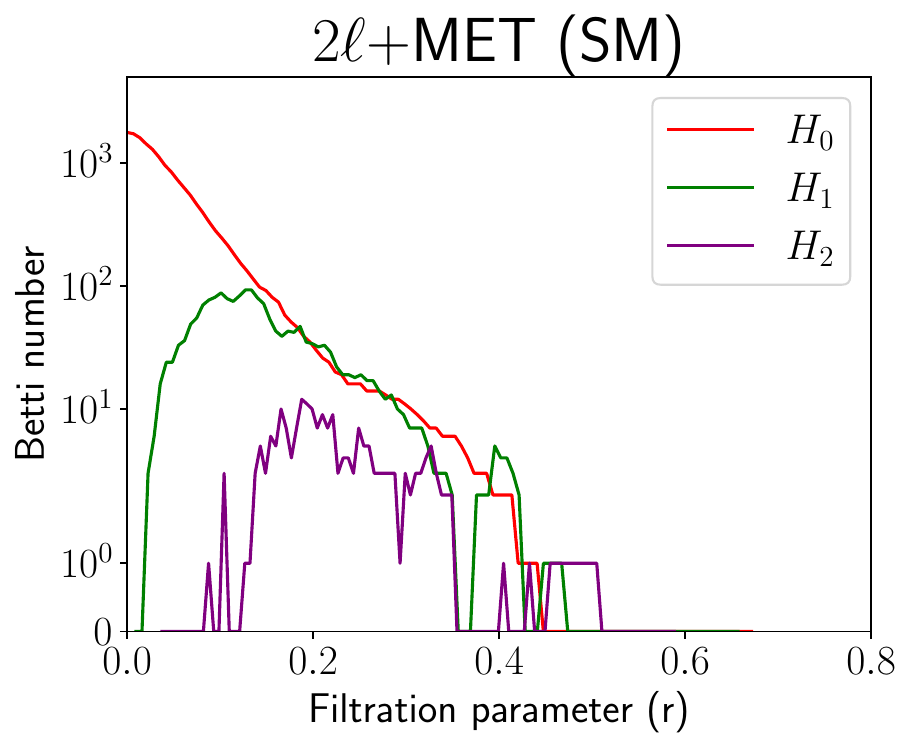}} 
    \subfloat[]{\includegraphics[width=0.3\textwidth]{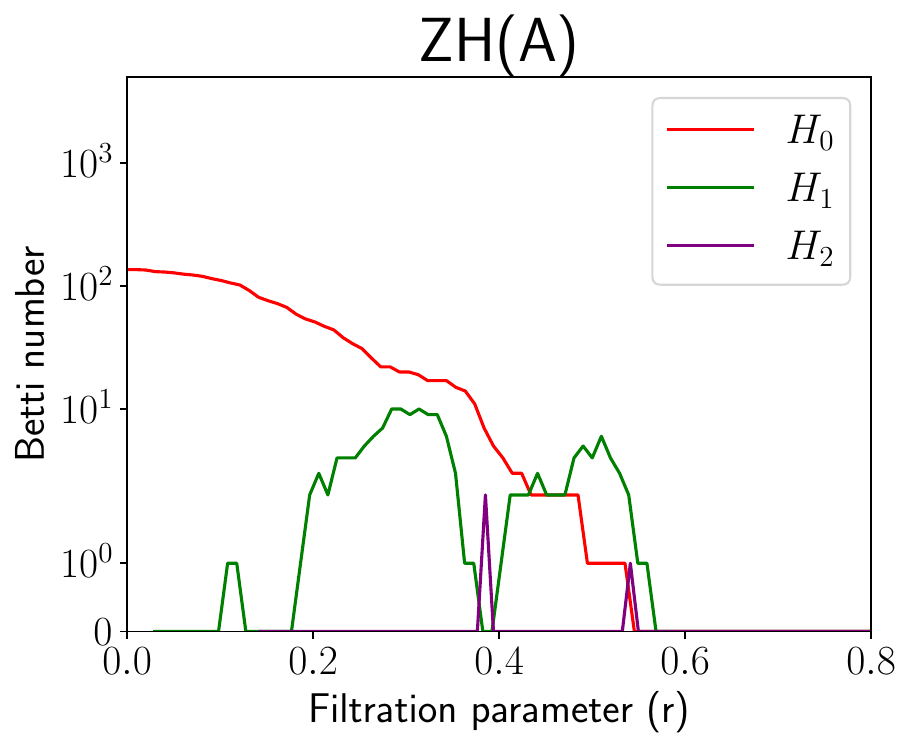}} 
    \subfloat[]{\includegraphics[width=0.3\textwidth]{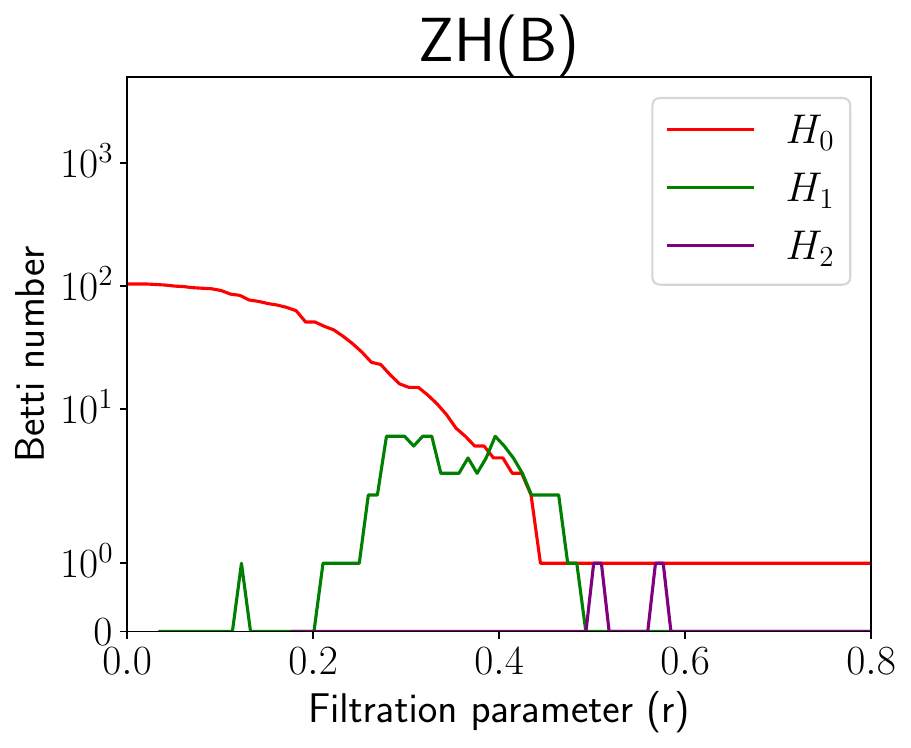}} 
    \caption{Betti curves for the Higgs invisible decay in the SM and the SSDM benchmark scenarios are presented. The first (second) row corresponds to WH (ZH) production mode.}
    \label{fig:ba_ssdm}
\end{figure*}
We have implemented this model in \texttt{SARAH v4.5.1} \cite{Staub:2013tta,Staub:2015kfa} and the spectra are obtained using \texttt{SARAH} generated \texttt{SPheno v4.0.5} \cite{Porod:2003um,Porod:2011nf} While keeping $\lambda_H=0.255$ to ensure the SM Higgs boson with $m_h\approx 125$ GeV and a low value for singlet quartic coupling strength $\lambda_s=0.1$, we have performed a scan over parameter space with
\begin{equation}
    |M_S^2|\leq 3\times10^3 \mathrm{GeV}^2, \;
    |\lambda_{sh}|\leq 0.1.
\end{equation}

\begin{table}[!ht]
\centering
\begin{tabular}{|c|c|c|c|c|c|} 
 \hline 
  Spectrum & $M_s^2$ & $\lambda_{sh}$ & $m_s$ (GeV) & $BR_{h\rightarrow s\:s}$ & Status   \\ \hline
  A &   -280	& 0.005	&  4.8  &  0.11 & Allowed\\ \hline
  B & 2235	& 0.005 &  50.0 &  0.07 & Allowed \\ \hline
  % C & -500	& 0.05 &  50.0 &  0.88 & Excluded \\ \hline
\end{tabular}
\caption{A and B are two benchmark scenarios in the SSDM having low and high singlet DM mass, respectively.}
\label{tab:ssdm_benchmarks}
\end{table}

In figure \ref{fig:scan_ssdm}(a) and \ref{fig:scan_ssdm}(b), we see a sharp edge ($m_s \approx 0$) with a slope of $135$ degree corresponding to $m_s^2 \geq 0$ in equation \ref{eq:ms_relation}. We also subject the parameter space under Higgs precision tests implemented by \texttt{HiggsTools v1.1.3} \cite{Bahl:2022igd} which include three sub-packages \texttt{HiggsPredictions}, \texttt{HiggsBounds} \cite{Bahl:2021yhk}, and \texttt{HiggsSignals} \cite{Bechtle:2020uwn} covering all publicly accessible experimental findings from 125 GeV Higgs boson measurements and searches for new (scalar) particles. The allowed regions are shaded in green and excluded regions in red in \ref{fig:scan_ssdm}(a) and \ref{fig:scan_ssdm}(c). We observe that only low values of $\lambda_{sh}$ are allowed under the Higgs precision test. This is kind of expected since large $\lambda_{sh}$ corresponds to large $h\rightarrow s \; s $ invisible decay. This is corroborated in figure \ref{fig:scan_ssdm}(d) with low BR($h\rightarrow s \; s $) for low values of $\lambda_{sh}$.

Based on the allowed regions of parameter space, we choose two benchmark scenarios (table \ref{tab:ssdm_benchmarks}) for demonstrating the global topological features associated with the signals involving invisible decay of Higgs boson to the real singlet DM in the SSDM. We also will contrast the findings with the SM counterpart. 

\section{Persistent homology of invisible Higgs decays in the SM and the SSDM}
\label{sec:ph-ssdm}

In the SM, extremely low (0.1$\%$) branching fraction to invisible final states is primarily through $H \rightarrow Z Z^{(*)} \rightarrow \nu \bar{\nu}\nu \bar{\nu}$ \cite{aad2022search,tumasyan2022search}. On the contrary, many BSM models feature a sizable amount of Higgs invisible decay to massive stable particles protected under some global symmetries, as in the SSDM mentioned previously. Thus, we expect a characteristic difference between the global topological properties of the events associated with the BSM model and the SM, particularly in the invisible decay of Higgs.

Table \ref{tab:ssdm_benchmarks} presents two benchmark scenarios allowed under Higgs precision tests. Benchmark-A and benchmark-B have similar coupling strengths ($\lambda_{sh}=0.005$) between singlet scalar and Higgs boson. However, benchmark-A has $m_s=4.8$ GeV and benchmark-B has $m_s=50$ GeV. Thus, the BR($h\rightarrow s \; s $) for benchmark-B is lower than benchmark-A due to compression in available phase space for decay.

For the SSDM case, we consider resonant production of ZH and WH with leptonic decay of the Z/W boson and the invisible decay of H ($H\rightarrow s\:s $). For background estimation, we consider electroweak resonant productions with leptonic decay of the Z/W boson and invisible decay of Z and H in the SM. This leads to two kinds of signals, i.e., $1\ell+MET$ and $2\ell+MET$ for benchmark-A and benchmark-B. We perform collider simulation upto detector level similar to section \ref{sec:framework}.

We apply $p_T^{l}>[50,40]$ GeV depending on the lepton counts. For $Z\rightarrow l^+ l^-$ process, we keep the leptonic invariant mass window from 80 GeV to 100 GeV. All the processes, including the SM background processes, we keep leptonic transverse mass $m_T^{l_1} < 40$ GeV and $MET<150$ GeV to effectively filter the SM backgrounds.

The $birth$-$death$ chart or the persistent diagram for the SM backgrounds and the benchmark scenarios of the SSDM are compared in figure \ref{fig:pe_ssdm}. In table \ref{tab:pe_ssdm}, we give persistent entropies associated with the persistent diagrams in figure \ref{fig:pe_ssdm} for up to the second homology dimension. We see figure \ref{fig:pe_ssdm} (e) and (f) are sparsely populated compared to other plots. This can be attributed to very low resonant production cross sections of ZH for benchmark-A (4.4 fb) and benchmark-B (2.8 fb), leading to lesser points at a fixed integrated luminosity. 
 
\begin{table}[!h]
\centering
\begin{tabular}{|l |c|c |c|} 
 \hline 
  Event type & $S_0$ & $S_1$ & $S_2$  \\ \hline
  $1\ell$+MET (SM) & 1.52	& 2.28 & 5.44\\ \hline
  WH (A) &  1.53 & 2.85	& -4.92 \\ \hline
  WH (B) & 1.52	& 3.17	& -4.60 \\ \hline
  $2\ell$+MET (SM)  &  1.56	& 2.61	& -10.51 \\ \hline
  ZH (A) &  1.47 & 29.72 & -0.37 \\ \hline
  ZH (B) & 1.50	& -2.98	& -0.19 \\ \hline
\end{tabular}
\caption{Persistent entropy associated with the Higgs invisible decay in the SM and the benchmark scenarios of the SSDM.}
\label{tab:pe_ssdm}
\end{table}

Referring to the table \ref{tab:pe_ssdm}, we find that in the case of the SM, the persistent entropy $S_1$ is slightly lower compared to the SSDM in both WH and ZH modes. On the contrary, $S_2$ is larger in magnitudes for the SM than for the SSDM. This can be corroborated by a larger spread of purple triangles above the diagonal in the SM compared to the SSDM.

\begin{table}[!h]
\centering
\begin{tabular}{|l |c|c |c|c|} 
 \hline 
  Event type & $BA_0$ & $BA_1$ & $BA_2$ \\ \hline
  $1\ell+MET$ (SM) & 165.08 (0.87) & 20.09 (0.11)	& 3.31 (0.01)\\ \hline
  WH (A) & 67.44 (0.91) & 5.94	(0.08) & 0.35 (0.00)\\ \hline
  WH (B) & 48.52 (0.93) & 3.63	(0.07) & 0.15 (0.00)\\ \hline
  $2\ell+MET$ (SM) & 103.35 (0.89) & 11.92 (0.10) & 1.16 (0.01)\\ \hline
  ZH (A) & 25.61 (0.94) & 1.51 (0.06) &	0.03 (0.00)\\ \hline
  ZH (B) & 21.94 (0.96) & 0.91 (0.04)	& 0.03 (0.00)\\ \hline
\end{tabular}
\caption{Betti areas associated with the Higgs invisible decay in the SM and the benchmark scenarios of the SSDM. Fractional Betti areas are shown in brackets.}
\label{tab:ba_ssdm}
\end{table}

In figure \ref{fig:ba_ssdm}, we present Betti curves for the SM production modes, along with the SSDM benchmark scenarios. The corresponding Betti areas are shown in table \ref{tab:ba_ssdm}. We see that zeroth Betti curves for ZH production mode in both benchmark scenarios drop slower than WH mode. Also, the Betti curves corresponding to $H_1$ for ZH mode peaks at a larger $r\approx 0.3$ compared to WH mode. However, Betti areas for both WH and ZH are smaller compared to the SM background. This is due to the lower production cross sections of the WH and ZH modes in the SSDM. Interestingly, the fractional Betti areas corresponding to $H_0$ for both benchmark-A and benchmark-B are larger compared to the SM background. On the other hand, we see the opposite trend for the Betti areas corresponding to $H_{1,2}$.

% Unlike the persistent entropies listed in table \ref{tab:pe_ssdm}, both WH(SM) and ZH(SM)  scenarios offer lower Betti areas compared to the SSDM benchmark scenarios across all homology dimensions. On the other hand, benchmark-A and benchmark-B also differ across all homology dimensions in WH and ZH scenarios. Benchmark-A with lower singlet DM mass (4.8 GeV) has the largest Betti areas for all homology dimensions in WH and ZH modes. We also observe that the zeroth Betti curve for the WH (ZH) mode of the SM drops to zero at a larger (shorter) filtration parameter than the SSDM. For benchmark-A and benchmark-B, the shape of the Betti curves for the second homology group ($H_2$) is characteristically different from the SM case.

\section{Summary and Conclusion}
\label{sec:conclusion}
Observations recorded by several detectors at LHC involve a complex interplay of countless variables in action. In such a complex experimental setup, discovering several fundamental particles predicted by the quantum field theory models is triumphant for particle physics. In this preliminary study, we suggest a generic novel framework to investigate the properties of the ensemble of event space as a whole using persistent homology. This technique is slowly gaining popularity among the data science community as complementary to classical machine learning approaches.

Philosophically, such global topological properties also establish that the system as a whole is as characteristically important as its components. We suggest that the signature of fundamental laws of Nature may also be found in the very complex global relations and information geometry associated with the LHC observations. This work serves as an exploratory step in that direction.

We have demonstrated the usability of the framework first for the SM electroweak sector and subsequently for the Higgs invisible decays in the SM and the SSDM. We find the characteristic difference between different physics scenarios encoded in the persistent diagrams and Betti curves for different homology dimensions. The associated persistent entropies and Betti areas also serve as the topological markers for the processes under discussion. We find that these global topological properties are useful properties that can supplement and complement the kinematic variables used for event-by-event in signal and background discrimination. These features can also be used with machine learning for discriminating the new physics scenarios from the SM background. 

\section*{Declaration of competing interest}
The authors declare that they have no known competing financial interests or personal relationships that could have appeared to influence the work reported in this paper.
\section*{Acknowledgements}
JB thanks IKSMHA Centre, IIT Mandi and IKS Centre, ISS Delhi for their support while part of the work was completed.

% %% The Appendices part is started with the command \appendix;
% %% appendix sections are then done as normal sections
% \appendix

% \section{Appendix title 1}
% %% \label{}

% \section{Appendix title 2}
% %% \label{}

\bibliographystyle{elsarticle-num-names} 
\bibliography{tda}

\begin{thebibliography}{54}
\expandafter\ifx\csname natexlab\endcsname\relax\def\natexlab#1{#1}\fi
\providecommand{\url}[1]{\texttt{#1}}
\providecommand{\href}[2]{#2}
\providecommand{\path}[1]{#1}
\providecommand{\DOIprefix}{doi:}
\providecommand{\ArXivprefix}{arXiv:}
\providecommand{\URLprefix}{URL: }
\providecommand{\Pubmedprefix}{pmid:}
\providecommand{\doi}[1]{\href{http://dx.doi.org/#1}{\path{#1}}}
\providecommand{\Pubmed}[1]{\href{pmid:#1}{\path{#1}}}
\providecommand{\bibinfo}[2]{#2}
\ifx\xfnm\relax \def\xfnm[#1]{\unskip,\space#1}\fi
%Type = Article
\bibitem[{Aad et~al.(2012)Aad, Abajyan, Abbott, Abdallah, Khalek, Abdelalim,
  Aben, Abi, Abolins, AbouZeid et~al.}]{aad2012observation}
\bibinfo{author}{G.~Aad}, \bibinfo{author}{T.~Abajyan},
  \bibinfo{author}{B.~Abbott}, \bibinfo{author}{J.~Abdallah},
  \bibinfo{author}{S.~A. Khalek}, \bibinfo{author}{A.~A. Abdelalim},
  \bibinfo{author}{R.~Aben}, \bibinfo{author}{B.~Abi},
  \bibinfo{author}{M.~Abolins}, \bibinfo{author}{O.~AbouZeid}, et~al.,
\newblock \bibinfo{title}{Observation of a new particle in the search for the
  standard model higgs boson with the atlas detector at the lhc},
\newblock \bibinfo{journal}{Physics Letters B} \bibinfo{volume}{716}
  (\bibinfo{year}{2012}) \bibinfo{pages}{1--29}.
%Type = Article
\bibitem[{Chatrchyan et~al.(2012)Chatrchyan, Khachatryan, Sirunyan, Tumasyan,
  Adam, Aguilo, Bergauer, Dragicevic, Er{\"o}, Fabjan
  et~al.}]{chatrchyan2012observation}
\bibinfo{author}{S.~Chatrchyan}, \bibinfo{author}{V.~Khachatryan},
  \bibinfo{author}{A.~M. Sirunyan}, \bibinfo{author}{A.~Tumasyan},
  \bibinfo{author}{W.~Adam}, \bibinfo{author}{E.~Aguilo},
  \bibinfo{author}{T.~Bergauer}, \bibinfo{author}{M.~Dragicevic},
  \bibinfo{author}{J.~Er{\"o}}, \bibinfo{author}{C.~Fabjan}, et~al.,
\newblock \bibinfo{title}{Observation of a new boson at a mass of 125 gev with
  the cms experiment at the lhc},
\newblock \bibinfo{journal}{Physics Letters B} \bibinfo{volume}{716}
  (\bibinfo{year}{2012}) \bibinfo{pages}{30--61}.
%Type = Article
\bibitem[{Franceschini et~al.(2022)Franceschini, Kim, Kong, Matchev, Park, and
  Shyamsundar}]{Franceschini:2022vck}
\bibinfo{author}{R.~Franceschini}, \bibinfo{author}{D.~Kim},
  \bibinfo{author}{K.~Kong}, \bibinfo{author}{K.~T. Matchev},
  \bibinfo{author}{M.~Park}, \bibinfo{author}{P.~Shyamsundar},
\newblock \bibinfo{title}{{Kinematic Variables and Feature Engineering for
  Particle Phenomenology}}  (\bibinfo{year}{2022}).
  \href{http://arxiv.org/abs/2206.13431}{{\tt arXiv:2206.13431}}.
%Type = Article
\bibitem[{Debnath et~al.(2016{\natexlab{a}})Debnath, Gainer, Kim, and
  Matchev}]{Debnath:2015wra}
\bibinfo{author}{D.~Debnath}, \bibinfo{author}{J.~S. Gainer},
  \bibinfo{author}{D.~Kim}, \bibinfo{author}{K.~T. Matchev},
\newblock \bibinfo{title}{{Edge Detecting New Physics the Voronoi Way}},
\newblock \bibinfo{journal}{EPL} \bibinfo{volume}{114}
  (\bibinfo{year}{2016}{\natexlab{a}}) \bibinfo{pages}{41001}.
  \DOIprefix\doi{10.1209/0295-5075/114/41001}.
  \href{http://arxiv.org/abs/1506.04141}{{\tt arXiv:1506.04141}}.
%Type = Article
\bibitem[{Debnath et~al.(2016{\natexlab{b}})Debnath, Gainer, Kilic, Kim,
  Matchev, and Yang}]{Debnath:2016mwb}
\bibinfo{author}{D.~Debnath}, \bibinfo{author}{J.~S. Gainer},
  \bibinfo{author}{C.~Kilic}, \bibinfo{author}{D.~Kim}, \bibinfo{author}{K.~T.
  Matchev}, \bibinfo{author}{Y.-P. Yang},
\newblock \bibinfo{title}{{Identifying Phase Space Boundaries with Voronoi
  Tessellations}},
\newblock \bibinfo{journal}{Eur. Phys. J. C} \bibinfo{volume}{76}
  (\bibinfo{year}{2016}{\natexlab{b}}) \bibinfo{pages}{645}.
  \DOIprefix\doi{10.1140/epjc/s10052-016-4431-z}.
  \href{http://arxiv.org/abs/1606.02721}{{\tt arXiv:1606.02721}}.
%Type = Article
\bibitem[{Matchev et~al.(2020)Matchev, Roman, and
  Shyamsundar}]{Matchev:2020vhr}
\bibinfo{author}{K.~T. Matchev}, \bibinfo{author}{A.~Roman},
  \bibinfo{author}{P.~Shyamsundar},
\newblock \bibinfo{title}{{Finding wombling boundaries in LHC data with Voronoi
  and Delaunay tessellations}},
\newblock \bibinfo{journal}{JHEP} \bibinfo{volume}{12} (\bibinfo{year}{2020})
  \bibinfo{pages}{137}. \DOIprefix\doi{10.1007/JHEP12(2020)137}.
  \href{http://arxiv.org/abs/2006.06582}{{\tt arXiv:2006.06582}}.
%Type = Article
\bibitem[{Mullin et~al.(2021)Mullin, Nicholls, Pacey, Parker, White, and
  Williams}]{Mullin:2019mmh}
\bibinfo{author}{A.~Mullin}, \bibinfo{author}{S.~Nicholls},
  \bibinfo{author}{H.~Pacey}, \bibinfo{author}{M.~Parker},
  \bibinfo{author}{M.~White}, \bibinfo{author}{S.~Williams},
\newblock \bibinfo{title}{{Does SUSY have friends? A new approach for LHC event
  analysis}},
\newblock \bibinfo{journal}{JHEP} \bibinfo{volume}{02} (\bibinfo{year}{2021})
  \bibinfo{pages}{160}. \DOIprefix\doi{10.1007/JHEP02(2021)160}.
  \href{http://arxiv.org/abs/1912.10625}{{\tt arXiv:1912.10625}}.
%Type = Article
\bibitem[{Nachman and Thaler(2021)}]{Nachman:2021yvi}
\bibinfo{author}{B.~Nachman}, \bibinfo{author}{J.~Thaler},
\newblock \bibinfo{title}{{Learning from many collider events at once}},
\newblock \bibinfo{journal}{Phys. Rev. D} \bibinfo{volume}{103}
  (\bibinfo{year}{2021}) \bibinfo{pages}{116013}.
  \DOIprefix\doi{10.1103/PhysRevD.103.116013}.
  \href{http://arxiv.org/abs/2101.07263}{{\tt arXiv:2101.07263}}.
%Type = Article
\bibitem[{Taylor et~al.(2015)Taylor, Klimm, Harrington, Kram{\'a}r, Mischaikow,
  Porter, and Mucha}]{taylor2015topological}
\bibinfo{author}{D.~Taylor}, \bibinfo{author}{F.~Klimm}, \bibinfo{author}{H.~A.
  Harrington}, \bibinfo{author}{M.~Kram{\'a}r},
  \bibinfo{author}{K.~Mischaikow}, \bibinfo{author}{M.~A. Porter},
  \bibinfo{author}{P.~J. Mucha},
\newblock \bibinfo{title}{Topological data analysis of contagion maps for
  examining spreading processes on networks},
\newblock \bibinfo{journal}{Nature communications} \bibinfo{volume}{6}
  (\bibinfo{year}{2015}) \bibinfo{pages}{7723}.
%Type = Article
\bibitem[{Topaz et~al.(2015)Topaz, Ziegelmeier, and
  Halverson}]{topaz2015topological}
\bibinfo{author}{C.~M. Topaz}, \bibinfo{author}{L.~Ziegelmeier},
  \bibinfo{author}{T.~Halverson},
\newblock \bibinfo{title}{Topological data analysis of biological aggregation
  models},
\newblock \bibinfo{journal}{PloS one} \bibinfo{volume}{10}
  (\bibinfo{year}{2015}) \bibinfo{pages}{e0126383}.
%Type = Article
\bibitem[{Lloyd et~al.(2016)Lloyd, Garnerone, and Zanardi}]{lloyd2016quantum}
\bibinfo{author}{S.~Lloyd}, \bibinfo{author}{S.~Garnerone},
  \bibinfo{author}{P.~Zanardi},
\newblock \bibinfo{title}{Quantum algorithms for topological and geometric
  analysis of data},
\newblock \bibinfo{journal}{Nature communications} \bibinfo{volume}{7}
  (\bibinfo{year}{2016}) \bibinfo{pages}{10138}.
%Type = Article
\bibitem[{Gidea and Katz(2018)}]{gidea2018topological}
\bibinfo{author}{M.~Gidea}, \bibinfo{author}{Y.~Katz},
\newblock \bibinfo{title}{Topological data analysis of financial time series:
  Landscapes of crashes},
\newblock \bibinfo{journal}{Physica A: Statistical Mechanics and its
  Applications} \bibinfo{volume}{491} (\bibinfo{year}{2018})
  \bibinfo{pages}{820--834}.
%Type = Article
\bibitem[{Saggar et~al.(2018)Saggar, Sporns, Gonzalez-Castillo, Bandettini,
  Carlsson, Glover, and Reiss}]{saggar2018towards}
\bibinfo{author}{M.~Saggar}, \bibinfo{author}{O.~Sporns},
  \bibinfo{author}{J.~Gonzalez-Castillo}, \bibinfo{author}{P.~A. Bandettini},
  \bibinfo{author}{G.~Carlsson}, \bibinfo{author}{G.~Glover},
  \bibinfo{author}{A.~L. Reiss},
\newblock \bibinfo{title}{Towards a new approach to reveal dynamical
  organization of the brain using topological data analysis},
\newblock \bibinfo{journal}{Nature communications} \bibinfo{volume}{9}
  (\bibinfo{year}{2018}) \bibinfo{pages}{1399}.
%Type = Article
\bibitem[{Sizemore et~al.(2019)Sizemore, Phillips-Cremins, Ghrist, and
  Bassett}]{sizemore2019importance}
\bibinfo{author}{A.~E. Sizemore}, \bibinfo{author}{J.~E. Phillips-Cremins},
  \bibinfo{author}{R.~Ghrist}, \bibinfo{author}{D.~S. Bassett},
\newblock \bibinfo{title}{The importance of the whole: topological data
  analysis for the network neuroscientist},
\newblock \bibinfo{journal}{Network Neuroscience} \bibinfo{volume}{3}
  (\bibinfo{year}{2019}) \bibinfo{pages}{656--673}.
%Type = Article
\bibitem[{Murugan and Robertson(2019)}]{Murugan:2019alb}
\bibinfo{author}{J.~Murugan}, \bibinfo{author}{D.~Robertson},
\newblock \bibinfo{title}{{An Introduction to Topological Data Analysis for
  Physicists: From LGM to FRBs}}  (\bibinfo{year}{2019}).
  \href{http://arxiv.org/abs/1904.11044}{{\tt arXiv:1904.11044}}.
%Type = Article
\bibitem[{Cole and Shiu(2019)}]{cole2019topological}
\bibinfo{author}{A.~Cole}, \bibinfo{author}{G.~Shiu},
\newblock \bibinfo{title}{Topological data analysis for the string landscape},
\newblock \bibinfo{journal}{Journal of High Energy Physics}
  \bibinfo{volume}{2019} (\bibinfo{year}{2019}) \bibinfo{pages}{1--31}.
%Type = Article
\bibitem[{Chazal and Michel(2021)}]{chazal2021introduction}
\bibinfo{author}{F.~Chazal}, \bibinfo{author}{B.~Michel},
\newblock \bibinfo{title}{An introduction to topological data analysis:
  fundamental and practical aspects for data scientists},
\newblock \bibinfo{journal}{Frontiers in artificial intelligence}
  \bibinfo{volume}{4} (\bibinfo{year}{2021}) \bibinfo{pages}{108}.
%Type = Article
\bibitem[{Csaki(1996)}]{csaki1996minimal}
\bibinfo{author}{C.~Csaki},
\newblock \bibinfo{title}{The minimal supersymmetric standard model},
\newblock \bibinfo{journal}{Modern Physics Letters A} \bibinfo{volume}{11}
  (\bibinfo{year}{1996}) \bibinfo{pages}{599--613}.
%Type = Incollection
\bibitem[{Martin(1998)}]{martin1998supersymmetry}
\bibinfo{author}{S.~P. Martin},
\newblock \bibinfo{title}{A supersymmetry primer},
\newblock in: \bibinfo{booktitle}{Perspectives on supersymmetry},
  \bibinfo{publisher}{World Scientific}, \bibinfo{year}{1998}, pp.
  \bibinfo{pages}{1--98}.
%Type = Book
\bibitem[{Baer and Tata(2006)}]{Baer:2006rs}
\bibinfo{author}{H.~Baer}, \bibinfo{author}{X.~Tata}, \bibinfo{title}{{Weak
  scale supersymmetry: From superfields to scattering events}},
  \bibinfo{publisher}{Cambridge University Press}, \bibinfo{year}{2006}.
%Type = Article
\bibitem[{Ellwanger et~al.(2010)Ellwanger, Hugonie, and
  Teixeira}]{ellwanger2010next}
\bibinfo{author}{U.~Ellwanger}, \bibinfo{author}{C.~Hugonie},
  \bibinfo{author}{A.~M. Teixeira},
\newblock \bibinfo{title}{The next-to-minimal supersymmetric standard model},
\newblock \bibinfo{journal}{Physics Reports} \bibinfo{volume}{496}
  (\bibinfo{year}{2010}) \bibinfo{pages}{1--77}.
%Type = Article
\bibitem[{Cembranos et~al.(2007)Cembranos, Feng, and
  Strigari}]{cembranos2007exotic}
\bibinfo{author}{J.~A. Cembranos}, \bibinfo{author}{J.~L. Feng},
  \bibinfo{author}{L.~E. Strigari},
\newblock \bibinfo{title}{Exotic collider signals from the complete phase
  diagram of minimal universal extra dimensions},
\newblock \bibinfo{journal}{Physical Review D} \bibinfo{volume}{75}
  (\bibinfo{year}{2007}) \bibinfo{pages}{036004}.
%Type = Article
\bibitem[{Datta et~al.(2010)Datta, Kong, and Matchev}]{datta2010minimal}
\bibinfo{author}{A.~Datta}, \bibinfo{author}{K.~Kong}, \bibinfo{author}{K.~T.
  Matchev},
\newblock \bibinfo{title}{Minimal universal extra dimensions in
  calchep/comphep},
\newblock \bibinfo{journal}{New Journal of Physics} \bibinfo{volume}{12}
  (\bibinfo{year}{2010}) \bibinfo{pages}{075017}.
%Type = Article
\bibitem[{Ham et~al.(2005)Ham, Jeong, and Oh}]{ham2005electroweak}
\bibinfo{author}{S.~Ham}, \bibinfo{author}{Y.~Jeong}, \bibinfo{author}{S.~Oh},
\newblock \bibinfo{title}{Electroweak phase transition in an extension of the
  standard model with a real higgs singlet},
\newblock \bibinfo{journal}{Journal of Physics G: Nuclear and Particle Physics}
  \bibinfo{volume}{31} (\bibinfo{year}{2005}) \bibinfo{pages}{857}.
%Type = Article
\bibitem[{Barger et~al.(2008)Barger, Langacker, McCaskey, Ramsey-Musolf, and
  Shaughnessy}]{barger2008cern}
\bibinfo{author}{V.~Barger}, \bibinfo{author}{P.~Langacker},
  \bibinfo{author}{M.~McCaskey}, \bibinfo{author}{M.~J. Ramsey-Musolf},
  \bibinfo{author}{G.~Shaughnessy},
\newblock \bibinfo{title}{Cern lhc phenomenology of an extended standard model
  with a real scalar singlet},
\newblock \bibinfo{journal}{Physical Review D} \bibinfo{volume}{77}
  (\bibinfo{year}{2008}) \bibinfo{pages}{035005}.
%Type = Article
\bibitem[{Guo and Wu(2010)}]{guo2010real}
\bibinfo{author}{W.-L. Guo}, \bibinfo{author}{Y.-L. Wu},
\newblock \bibinfo{title}{The real singlet scalar dark matter model},
\newblock \bibinfo{journal}{Journal of High Energy Physics}
  \bibinfo{volume}{2010} (\bibinfo{year}{2010}) \bibinfo{pages}{1--13}.
%Type = Article
\bibitem[{Enqvist et~al.(2014)Enqvist, Nurmi, Tenkanen, and
  Tuominen}]{enqvist2014standard}
\bibinfo{author}{K.~Enqvist}, \bibinfo{author}{S.~Nurmi},
  \bibinfo{author}{T.~Tenkanen}, \bibinfo{author}{K.~Tuominen},
\newblock \bibinfo{title}{Standard model with a real singlet scalar and
  inflation},
\newblock \bibinfo{journal}{Journal of Cosmology and Astroparticle Physics}
  \bibinfo{volume}{2014} (\bibinfo{year}{2014}) \bibinfo{pages}{035}.
%Type = Article
\bibitem[{Feng et~al.(2015)Feng, Profumo, and Ubaldi}]{feng2015closing}
\bibinfo{author}{L.~Feng}, \bibinfo{author}{S.~Profumo},
  \bibinfo{author}{L.~Ubaldi},
\newblock \bibinfo{title}{Closing in on singlet scalar dark matter: Lux,
  invisible higgs decays and gamma-ray lines},
\newblock \bibinfo{journal}{Journal of High Energy Physics}
  \bibinfo{volume}{2015} (\bibinfo{year}{2015}) \bibinfo{pages}{1--13}.
%Type = Article
\bibitem[{Kanemura et~al.(2016)Kanemura, Kikuchi, and
  Yagyu}]{kanemura2016radiative}
\bibinfo{author}{S.~Kanemura}, \bibinfo{author}{M.~Kikuchi},
  \bibinfo{author}{K.~Yagyu},
\newblock \bibinfo{title}{Radiative corrections to the higgs boson couplings in
  the model with an additional real singlet scalar field},
\newblock \bibinfo{journal}{Nuclear Physics B} \bibinfo{volume}{907}
  (\bibinfo{year}{2016}) \bibinfo{pages}{286--322}.
%Type = Article
\bibitem[{Kurup and Perelstein(2017)}]{kurup2017dynamics}
\bibinfo{author}{G.~Kurup}, \bibinfo{author}{M.~Perelstein},
\newblock \bibinfo{title}{Dynamics of electroweak phase transition in
  singlet-scalar extension of the standard model},
\newblock \bibinfo{journal}{Physical Review D} \bibinfo{volume}{96}
  (\bibinfo{year}{2017}) \bibinfo{pages}{015036}.
%Type = Article
\bibitem[{Chiang et~al.(2019)Chiang, Li, and Senaha}]{chiang2019revisiting}
\bibinfo{author}{C.-W. Chiang}, \bibinfo{author}{Y.-T. Li},
  \bibinfo{author}{E.~Senaha},
\newblock \bibinfo{title}{Revisiting electroweak phase transition in the
  standard model with a real singlet scalar},
\newblock \bibinfo{journal}{Physics Letters B} \bibinfo{volume}{789}
  (\bibinfo{year}{2019}) \bibinfo{pages}{154--159}.
%Type = Article
\bibitem[{Carlsson(2009)}]{carlsson2009topology}
\bibinfo{author}{G.~Carlsson},
\newblock \bibinfo{title}{Topology and data},
\newblock \bibinfo{journal}{Bulletin of the American Mathematical Society}
  \bibinfo{volume}{46} (\bibinfo{year}{2009}) \bibinfo{pages}{255--308}.
%Type = Misc
\bibitem[{giotto-tda library(2023)}]{Giotto:2023}
\bibinfo{author}{giotto-tda library}, \bibinfo{title}{Tutorials and examples},
  \bibinfo{year}{2023}. \URLprefix
  \url{https://giotto-ai.github.io/gtda-docs/0.5.1/notebooks}.
%Type = Misc
\bibitem[{library – Topological~data analysis and geometric inference
  in~higher dimensions(2023)}]{Gudhi:2023}
\bibinfo{author}{G.~library – Topological~data analysis},
  \bibinfo{author}{geometric inference in~higher dimensions},
  \bibinfo{title}{Tda-tutorial}, \bibinfo{year}{2023}. \URLprefix
  \url{https://github.com/GUDHI/TDA-tutorial}.
%Type = Inproceedings
\bibitem[{Takens(2006)}]{takens2006detecting}
\bibinfo{author}{F.~Takens},
\newblock \bibinfo{title}{Detecting strange attractors in turbulence},
\newblock in: \bibinfo{booktitle}{Dynamical Systems and Turbulence, Warwick
  1980: proceedings of a symposium held at the University of Warwick 1979/80},
  \bibinfo{organization}{Springer}, \bibinfo{year}{2006}, pp.
  \bibinfo{pages}{366--381}.
%Type = Article
\bibitem[{Vietoris(1927)}]{vietoris1927hoheren}
\bibinfo{author}{L.~Vietoris},
\newblock \bibinfo{title}{{\"U}ber den h{\"o}heren zusammenhang kompakter
  r{\"a}ume und eine klasse von zusammenhangstreuen abbildungen},
\newblock \bibinfo{journal}{Mathematische Annalen} \bibinfo{volume}{97}
  (\bibinfo{year}{1927}) \bibinfo{pages}{454--472}.
%Type = Article
\bibitem[{Alwall et~al.(2014)Alwall, Frederix, Frixione, Hirschi, Maltoni,
  Mattelaer, Shao, Stelzer, Torrielli, and Zaro}]{Alwall:2014hca}
\bibinfo{author}{J.~Alwall}, \bibinfo{author}{R.~Frederix},
  \bibinfo{author}{S.~Frixione}, \bibinfo{author}{V.~Hirschi},
  \bibinfo{author}{F.~Maltoni}, \bibinfo{author}{O.~Mattelaer},
  \bibinfo{author}{H.~S. Shao}, \bibinfo{author}{T.~Stelzer},
  \bibinfo{author}{P.~Torrielli}, \bibinfo{author}{M.~Zaro},
\newblock \bibinfo{title}{{The automated computation of tree-level and
  next-to-leading order differential cross sections, and their matching to
  parton shower simulations}},
\newblock \bibinfo{journal}{JHEP} \bibinfo{volume}{07} (\bibinfo{year}{2014})
  \bibinfo{pages}{079}. \DOIprefix\doi{10.1007/JHEP07(2014)079}.
  \href{http://arxiv.org/abs/1405.0301}{{\tt arXiv:1405.0301}}.
%Type = Article
\bibitem[{Frederix et~al.(2018)Frederix, Frixione, Hirschi, Pagani, Shao, and
  Zaro}]{Frederix:2018nkq}
\bibinfo{author}{R.~Frederix}, \bibinfo{author}{S.~Frixione},
  \bibinfo{author}{V.~Hirschi}, \bibinfo{author}{D.~Pagani},
  \bibinfo{author}{H.~S. Shao}, \bibinfo{author}{M.~Zaro},
\newblock \bibinfo{title}{{The automation of next-to-leading order electroweak
  calculations}},
\newblock \bibinfo{journal}{JHEP} \bibinfo{volume}{07} (\bibinfo{year}{2018})
  \bibinfo{pages}{185}. \DOIprefix\doi{10.1007/JHEP11(2021)085}.
  \href{http://arxiv.org/abs/1804.10017}{{\tt arXiv:1804.10017}},
  \bibinfo{note}{[Erratum: JHEP 11, 085 (2021)]}.
%Type = Article
\bibitem[{Ball et~al.(2017)Ball, Bertone, Carrazza, Debbio, Forte,
  Groth-Merrild, Guffanti, Hartland, Kassabov, Latorre et~al.}]{ball2017parton}
\bibinfo{author}{R.~D. Ball}, \bibinfo{author}{V.~Bertone},
  \bibinfo{author}{S.~Carrazza}, \bibinfo{author}{L.~D. Debbio},
  \bibinfo{author}{S.~Forte}, \bibinfo{author}{P.~Groth-Merrild},
  \bibinfo{author}{A.~Guffanti}, \bibinfo{author}{N.~P. Hartland},
  \bibinfo{author}{Z.~Kassabov}, \bibinfo{author}{J.~I. Latorre}, et~al.,
\newblock \bibinfo{title}{Parton distributions from high-precision collider
  data: Nnpdf collaboration},
\newblock \bibinfo{journal}{The European Physical Journal C}
  \bibinfo{volume}{77} (\bibinfo{year}{2017}) \bibinfo{pages}{1--75}.
%Type = Article
\bibitem[{Bierlich et~al.(2022)}]{Bierlich:2022pfr}
\bibinfo{author}{C.~Bierlich}, et~al.,
\newblock \bibinfo{title}{{A comprehensive guide to the physics and usage of
  PYTHIA 8.3}}  (\bibinfo{year}{2022}).
  \DOIprefix\doi{10.21468/SciPostPhysCodeb.8}.
  \href{http://arxiv.org/abs/2203.11601}{{\tt arXiv:2203.11601}}.
%Type = Article
\bibitem[{Cacciari and Salam(2006)}]{Cacciari:2005hq}
\bibinfo{author}{M.~Cacciari}, \bibinfo{author}{G.~P. Salam},
\newblock \bibinfo{title}{{Dispelling the $N^{3}$ myth for the $k_t$
  jet-finder}},
\newblock \bibinfo{journal}{Phys. Lett. B} \bibinfo{volume}{641}
  (\bibinfo{year}{2006}) \bibinfo{pages}{57--61}.
  \DOIprefix\doi{10.1016/j.physletb.2006.08.037}.
  \href{http://arxiv.org/abs/hep-ph/0512210}{{\tt arXiv:hep-ph/0512210}}.
%Type = Article
\bibitem[{Cacciari et~al.(2012)Cacciari, Salam, and Soyez}]{Cacciari:2011ma}
\bibinfo{author}{M.~Cacciari}, \bibinfo{author}{G.~P. Salam},
  \bibinfo{author}{G.~Soyez},
\newblock \bibinfo{title}{{FastJet User Manual}},
\newblock \bibinfo{journal}{Eur. Phys. J. C} \bibinfo{volume}{72}
  (\bibinfo{year}{2012}) \bibinfo{pages}{1896}.
  \DOIprefix\doi{10.1140/epjc/s10052-012-1896-2}.
  \href{http://arxiv.org/abs/1111.6097}{{\tt arXiv:1111.6097}}.
%Type = Article
\bibitem[{de~Favereau et~al.(2014)de~Favereau, Delaere, Demin, Giammanco,
  Lema\^\i{}tre, Mertens, and Selvaggi}]{deFavereau:2013fsa}
\bibinfo{author}{J.~de~Favereau}, \bibinfo{author}{C.~Delaere},
  \bibinfo{author}{P.~Demin}, \bibinfo{author}{A.~Giammanco},
  \bibinfo{author}{V.~Lema\^\i{}tre}, \bibinfo{author}{A.~Mertens},
  \bibinfo{author}{M.~Selvaggi} (\bibinfo{collaboration}{DELPHES 3}),
\newblock \bibinfo{title}{{DELPHES 3, A modular framework for fast simulation
  of a generic collider experiment}},
\newblock \bibinfo{journal}{JHEP} \bibinfo{volume}{02} (\bibinfo{year}{2014})
  \bibinfo{pages}{057}. \DOIprefix\doi{10.1007/JHEP02(2014)057}.
  \href{http://arxiv.org/abs/1307.6346}{{\tt arXiv:1307.6346}}.
%Type = Article
\bibitem[{Brun and Rademakers(1997)}]{brun1997root}
\bibinfo{author}{R.~Brun}, \bibinfo{author}{F.~Rademakers},
\newblock \bibinfo{title}{Root—an object oriented data analysis framework},
\newblock \bibinfo{journal}{Nuclear instruments and methods in physics research
  section A: accelerators, spectrometers, detectors and associated equipment}
  \bibinfo{volume}{389} (\bibinfo{year}{1997}) \bibinfo{pages}{81--86}.
%Type = Article
\bibitem[{Tauzin et~al.(2021)Tauzin, Lupo, Tunstall, P\'{e}rez, Caorsi,
  Medina-Mardones, Dassatti, and Hess}]{giotto-tda}
\bibinfo{author}{G.~Tauzin}, \bibinfo{author}{U.~Lupo},
  \bibinfo{author}{L.~Tunstall}, \bibinfo{author}{J.~B. P\'{e}rez},
  \bibinfo{author}{M.~Caorsi}, \bibinfo{author}{A.~M. Medina-Mardones},
  \bibinfo{author}{A.~Dassatti}, \bibinfo{author}{K.~Hess},
\newblock \bibinfo{title}{giotto-tda: A topological data analysis toolkit for
  machine learning and data exploration},
\newblock \bibinfo{journal}{Journal of Machine Learning Research}
  \bibinfo{volume}{22} (\bibinfo{year}{2021}) \bibinfo{pages}{1--6}. \URLprefix
  \url{http://jmlr.org/papers/v22/20-325.html}.
%Type = Article
\bibitem[{Staub(2014)}]{Staub:2013tta}
\bibinfo{author}{F.~Staub},
\newblock \bibinfo{title}{{SARAH 4 : A tool for (not only SUSY) model
  builders}},
\newblock \bibinfo{journal}{Comput. Phys. Commun.} \bibinfo{volume}{185}
  (\bibinfo{year}{2014}) \bibinfo{pages}{1773--1790}.
  \DOIprefix\doi{10.1016/j.cpc.2014.02.018}.
  \href{http://arxiv.org/abs/1309.7223}{{\tt arXiv:1309.7223}}.
%Type = Article
\bibitem[{Staub(2015)}]{Staub:2015kfa}
\bibinfo{author}{F.~Staub},
\newblock \bibinfo{title}{{Exploring new models in all detail with SARAH}},
\newblock \bibinfo{journal}{Adv. High Energy Phys.} \bibinfo{volume}{2015}
  (\bibinfo{year}{2015}) \bibinfo{pages}{840780}.
  \DOIprefix\doi{10.1155/2015/840780}.
  \href{http://arxiv.org/abs/1503.04200}{{\tt arXiv:1503.04200}}.
%Type = Article
\bibitem[{Porod(2003)}]{Porod:2003um}
\bibinfo{author}{W.~Porod},
\newblock \bibinfo{title}{{SPheno, a program for calculating supersymmetric
  spectra, SUSY particle decays and SUSY particle production at e+ e-
  colliders}},
\newblock \bibinfo{journal}{Comput. Phys. Commun.} \bibinfo{volume}{153}
  (\bibinfo{year}{2003}) \bibinfo{pages}{275--315}.
  \DOIprefix\doi{10.1016/S0010-4655(03)00222-4}.
  \href{http://arxiv.org/abs/hep-ph/0301101}{{\tt arXiv:hep-ph/0301101}}.
%Type = Article
\bibitem[{Porod and Staub(2012)}]{Porod:2011nf}
\bibinfo{author}{W.~Porod}, \bibinfo{author}{F.~Staub},
\newblock \bibinfo{title}{{SPheno 3.1: Extensions including flavour, CP-phases
  and models beyond the MSSM}},
\newblock \bibinfo{journal}{Comput. Phys. Commun.} \bibinfo{volume}{183}
  (\bibinfo{year}{2012}) \bibinfo{pages}{2458--2469}.
  \DOIprefix\doi{10.1016/j.cpc.2012.05.021}.
  \href{http://arxiv.org/abs/1104.1573}{{\tt arXiv:1104.1573}}.
%Type = Article
\bibitem[{Bahl et~al.(2023)Bahl, Biek\"otter, Heinemeyer, Li, Paasch, Weiglein,
  and Wittbrodt}]{Bahl:2022igd}
\bibinfo{author}{H.~Bahl}, \bibinfo{author}{T.~Biek\"otter},
  \bibinfo{author}{S.~Heinemeyer}, \bibinfo{author}{C.~Li},
  \bibinfo{author}{S.~Paasch}, \bibinfo{author}{G.~Weiglein},
  \bibinfo{author}{J.~Wittbrodt},
\newblock \bibinfo{title}{{HiggsTools: BSM scalar phenomenology with new
  versions of HiggsBounds and HiggsSignals}},
\newblock \bibinfo{journal}{Comput. Phys. Commun.} \bibinfo{volume}{291}
  (\bibinfo{year}{2023}) \bibinfo{pages}{108803}.
  \DOIprefix\doi{10.1016/j.cpc.2023.108803}.
  \href{http://arxiv.org/abs/2210.09332}{{\tt arXiv:2210.09332}}.
%Type = Article
\bibitem[{Bahl et~al.(2022)Bahl, Lozano, Stefaniak, and
  Wittbrodt}]{Bahl:2021yhk}
\bibinfo{author}{H.~Bahl}, \bibinfo{author}{V.~M. Lozano},
  \bibinfo{author}{T.~Stefaniak}, \bibinfo{author}{J.~Wittbrodt},
\newblock \bibinfo{title}{{Testing exotic scalars with HiggsBounds}},
\newblock \bibinfo{journal}{Eur. Phys. J. C} \bibinfo{volume}{82}
  (\bibinfo{year}{2022}) \bibinfo{pages}{584}.
  \DOIprefix\doi{10.1140/epjc/s10052-022-10446-2}.
  \href{http://arxiv.org/abs/2109.10366}{{\tt arXiv:2109.10366}}.
%Type = Article
\bibitem[{Bechtle et~al.(2021)Bechtle, Heinemeyer, Klingl, Stefaniak, Weiglein,
  and Wittbrodt}]{Bechtle:2020uwn}
\bibinfo{author}{P.~Bechtle}, \bibinfo{author}{S.~Heinemeyer},
  \bibinfo{author}{T.~Klingl}, \bibinfo{author}{T.~Stefaniak},
  \bibinfo{author}{G.~Weiglein}, \bibinfo{author}{J.~Wittbrodt},
\newblock \bibinfo{title}{{HiggsSignals-2: Probing new physics with precision
  Higgs measurements in the LHC 13 TeV era}},
\newblock \bibinfo{journal}{Eur. Phys. J. C} \bibinfo{volume}{81}
  (\bibinfo{year}{2021}) \bibinfo{pages}{145}.
  \DOIprefix\doi{10.1140/epjc/s10052-021-08942-y}.
  \href{http://arxiv.org/abs/2012.09197}{{\tt arXiv:2012.09197}}.
%Type = Article
\bibitem[{Aad et~al.(2022)Aad, Abbott, Abbott, Abed~Abud, Abeling,
  Abhayasinghe, Abidi, Aboulhorma, Abramowicz, Abreu et~al.}]{aad2022search}
\bibinfo{author}{G.~Aad}, \bibinfo{author}{B.~Abbott}, \bibinfo{author}{D.~C.
  Abbott}, \bibinfo{author}{A.~Abed~Abud}, \bibinfo{author}{K.~Abeling},
  \bibinfo{author}{D.~K. Abhayasinghe}, \bibinfo{author}{S.~H. Abidi},
  \bibinfo{author}{A.~Aboulhorma}, \bibinfo{author}{H.~Abramowicz},
  \bibinfo{author}{H.~Abreu}, et~al.,
\newblock \bibinfo{title}{Search for invisible higgs-boson decays in events
  with vector-boson fusion signatures using 139 fb- 1 of proton-proton data
  recorded by the atlas experiment},
\newblock \bibinfo{journal}{Journal of High Energy Physics}
  \bibinfo{volume}{2022} (\bibinfo{year}{2022}) \bibinfo{pages}{1--66}.
%Type = Article
\bibitem[{Tumasyan et~al.(2022)Tumasyan, Adam, Andrejkovic, Bergauer,
  Chatterjee, Damanakis, Dragicevic, Del~Valle, Fruehwirth, Jeitler
  et~al.}]{tumasyan2022search}
\bibinfo{author}{A.~Tumasyan}, \bibinfo{author}{W.~Adam},
  \bibinfo{author}{J.~W. Andrejkovic}, \bibinfo{author}{T.~Bergauer},
  \bibinfo{author}{S.~Chatterjee}, \bibinfo{author}{K.~Damanakis},
  \bibinfo{author}{M.~Dragicevic}, \bibinfo{author}{A.~E. Del~Valle},
  \bibinfo{author}{R.~Fruehwirth}, \bibinfo{author}{M.~Jeitler}, et~al.,
\newblock \bibinfo{title}{Search for invisible decays of the higgs boson
  produced via vector boson fusion in proton-proton collisions at s= 13 tev},
\newblock \bibinfo{journal}{Physical Review D} \bibinfo{volume}{105}
  (\bibinfo{year}{2022}) \bibinfo{pages}{092007}.

\end{thebibliography}



%% For example:

%% \bibitem[Aladro et al.(2015)]{Aladro15} Aladro, R., Martín, S., Riquelme, D., et al. 2015, \aas, 579, A101

%%\end{thebibliography}

\end{document}